\documentclass[aps,prb,superscriptaddress,twocolumn]{revtex4-2}
\usepackage{graphics}
\usepackage{epsfig}
\usepackage{color}
\usepackage{stackrel}
\usepackage{amsfonts}
\usepackage{wrapfig}
\usepackage{pstricks}
\usepackage{pst-node}
\usepackage{bm}
\usepackage{dcolumn}
\usepackage{multirow}
\usepackage{epstopdf}
\usepackage{subfigure}
\usepackage{amssymb}
\usepackage{amsmath}
\usepackage{commath}
\usepackage{graphicx,bm}
\usepackage{verbatim}
\usepackage{booktabs}
\usepackage[para]{threeparttable}
\usepackage{etoolbox}
\usepackage{lipsum}

\begin{document}
\title{Relaxation and darkening of excitonic complexes in electrostatically-doped monolayer semiconductors: Roles of exciton-electron and trion-electron interactions}
%

\author{Min~Yang}
\altaffiliation{The first three authors have equal contributions}
\affiliation{Department of Electrical and Computer Engineering, University of Rochester, Rochester, New York 14627, USA}
\author{Lei Ren}
\altaffiliation{The first three authors have equal contributions}
\affiliation{Universit\'{e} de Toulouse, INSA-CNRS-UPS, LPCNO, 135 Av. Rangueil, 31077 Toulouse, France}
\author{Cedric Robert}
\altaffiliation{The first three authors have equal contributions}
\affiliation{Universit\'{e} de Toulouse, INSA-CNRS-UPS, LPCNO, 135 Av. Rangueil, 31077 Toulouse, France}
\author{Dinh~Van~Tuan}
\affiliation{Department of Electrical and Computer Engineering, University of Rochester, Rochester, New York 14627, USA}
\author{Laurent Lombez}
\affiliation{Universit\'{e} de Toulouse, INSA-CNRS-UPS, LPCNO, 135 Av. Rangueil, 31077 Toulouse, France}
\author{Bernhard Urbaszek}
\affiliation{Universit\'{e} de Toulouse, INSA-CNRS-UPS, LPCNO, 135 Av. Rangueil, 31077 Toulouse, France}%
\author{Xavier Marie}
\affiliation{Universit\'{e} de Toulouse, INSA-CNRS-UPS, LPCNO, 135 Av. Rangueil, 31077 Toulouse, France}
\author{Hanan~Dery}
\altaffiliation{hanan.dery@rochester.edu}
\affiliation{Department of Electrical and Computer Engineering, University of Rochester, Rochester, New York 14627, USA}
\affiliation{Department of Physics and Astronomy, University of Rochester, Rochester, New York 14627, USA}

\begin{abstract}
We present  photoluminescence measurements in monolayer WSe$_2$, which point to the importance of the interaction between charged particles and excitonic complexes. The theoretical analysis highlights the key role played by exchange scattering, referring to cases wherein the particle composition of the complex changes after the interaction. For example, exchange scattering renders bright excitonic complexes dark in monolayer  WSe$_2$ on accounts of the unique valley-spin configuration in this material.  In addition to the ultrafast energy relaxation of hot excitonic complexes following their interaction with electrons or holes, our analysis sheds light on several key features that are commonly seen in the photoluminescence of this monolayer semiconductor. In particular, we can understand why the photoluminescence intensity of the neutral bright exciton is strongest when the monolayer is hole-doped rather than charge neutral or electron-doped. Or the reason for the dramatic  increase of the photoluminescence intensity of negatively-charged excitons (trions) as soon as  electrons are added to the monolayer. To self-consistently explain the findings, we further study the photoluminescence spectra at different excitation energies and analyze the behavior of the elusive indirect exciton.
\end{abstract}

\maketitle

\section{Introduction}\label{sec:intro}

The improvement in fabrication of high-quality monolayer transition-metal dichalcogenides (ML-TMDs) devices in recent years have sharpened the understanding of optical properties in these materials \cite{Wang_RMP18,Mak_NatPhoton18}. This progress led to identification of various species of charged excitons (trions) \cite{Jones_NatPhys16,Courtade_PRB17}, dark excitons \cite{Zhang_NatNano17,Zhou_NatNano17,Wang_PRL17}, neutral and charged biexctions \cite{Chen_NatComm18,Ye_NatComm18,Li_NatComm18,Barbone_NatComm18}, indirect excitons and their phonon replicas \cite{He_NatComm20,Liu_PRL20}, as well as dark trions and their phonon replicas \cite{Li_ACS19,He_NatComm20,Liu_PRL20,Tang_NatComm19}. In addition to showing multitude of excitonic complexes, low-temperature photoluminescence (PL) experiments in ML-WSe$_2$ reveal non-trivial dependence on charge density \cite{Chen_NatComm18,Ye_NatComm18,Li_NatComm18,Barbone_NatComm18,He_NatComm20,Liu_PRL20}. To understand this dependence, one should know how off-resonance photoexcitation generates excitons in the presence of charge carriers, and how these excitons relax in energy and transform to other few-body complexes. In this work, we will focus on the relaxation and darkening of bright excitonic complexes during their relaxation. 

\begin{figure}
\includegraphics[width=8cm]{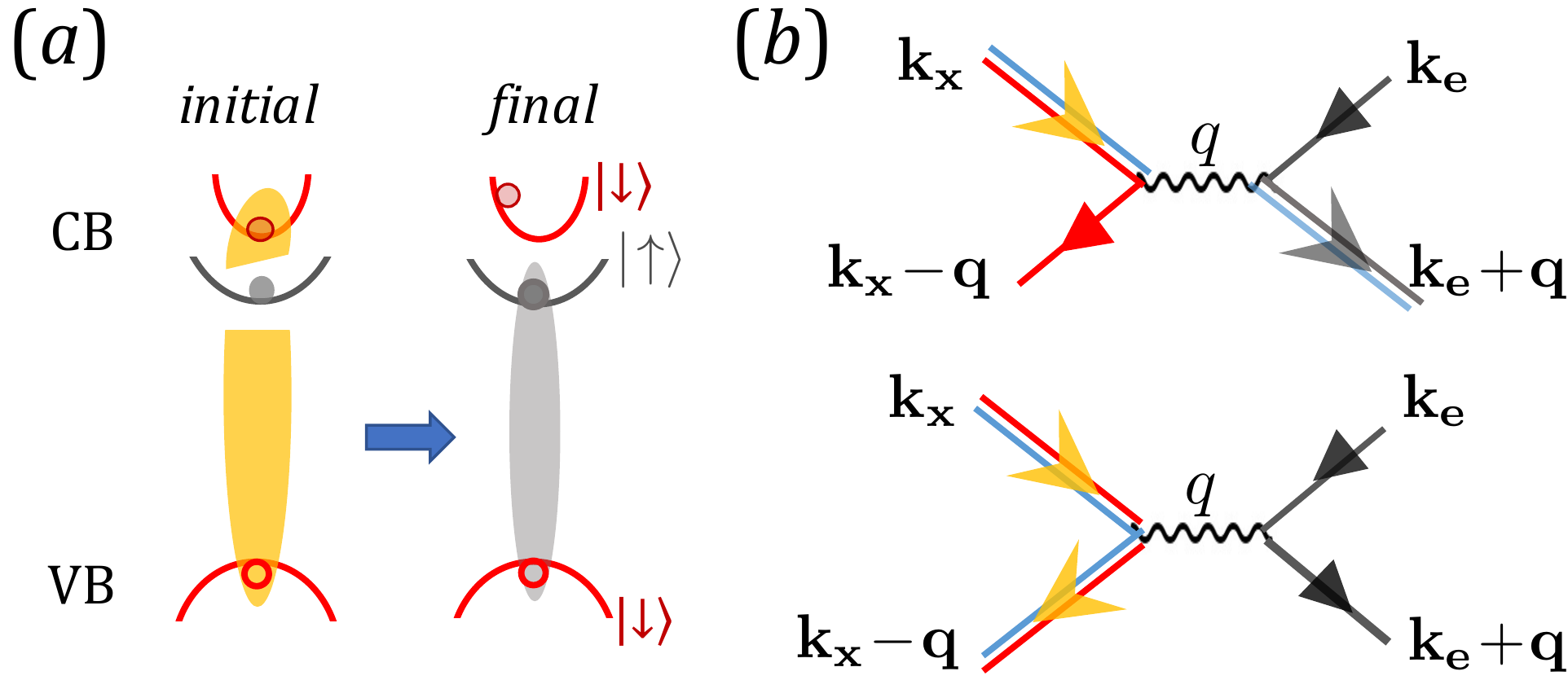}
 \caption{The exciton-electron scattering process. (a) An exciton, whose electron component belongs to the top valley of the CB, scatters with an incoming electron from the bottom CB valley. The electron of the exciton in the final state is exchanged with the incoming free electron. (b) Feynman diagrams of the scattering process. Double (single) lines denote bound excitons (free electrons). The upper diagram corresponds to exchange scattering, such as the one leading to the transition in (a). The lower diagram corresponds to direct scattering, in which the same electron-hole pair remains bound before and after scattering. As we will explain, the direct scattering amplitude is weak.} \label{fig:scheme}
\end{figure}

For an excitonic complex to be bright/dark, the orbital transition between one of its electrons and one of its holes has to be dipole allowed/forbidden, whereas the spins of the electron in the conduction band (CB) and the missing electron in the valence band (VB) have to be parallel/antiparallel \cite{Dery_PRB15,Slobodeniuk_2DMater16,Scharf_JPCM19}. Figure~\ref{fig:scheme}(a) shows an example for bright and spin-forbidden dark excitons where the spin of the energy bands is color coded. The bright exciton in this example has its electron in the top valley of the CB, and this choice will become clear later. When the bright exciton scatters off an incoming (thermal) electron in the bottom CB valley, exchange scattering refers to switching the electron of the exciton in the final state, as shown in the top Feynman diagram of Fig.~\ref{fig:scheme}(b). The result of this scattering is not only a means for ultrafast energy relaxation of hot excitons \cite{Honold_PRB89,Eccleston_PRB91}, but also for their darkening. The exchange scattering is relatively strong despite the neutrality of the exciton \cite{Kochereshko_PRB98}, whereas the direct scattering component is weak (i.e., when the exciton retains its original components, as shown in the bottom Feynman diagram of Fig.~\ref{fig:scheme}(b)) \cite{Ramon_PRB03,Shahnazaryan_PRB17}. It is noted that if the charge density is relatively small, the exciton-electron scattering is more probable than their binding to form a trion. The reason is that binding requires external agents to conserve both energy and momentum (e.g., phonons or other free electrons that can receive the gained energy from binding). 

The goal of this work is to establish connection between various phenomena seen in the PL of ML-WSe$_2$ and the interaction of excitonic complexes with electrons (or holes). One phenomenon, measured by several groups and shown in Fig.~\ref{fig:map}, is the enhanced PL intensity of bright excitons when the monolayer is electrostatically-doped with holes \cite{Ye_NatComm18,Barbone_NatComm18,Liu_PRL19, Li_NanoLett19}: The PL intensity of $X^0$ in Fig.~\ref{fig:map} is clearly larger when the gate voltage is negative  ($p$-doping) compared to positive  ($n$-doping). This observation is somewhat counterintuitive since the expectation is that the PL of excitons is strongest when the ML is charge neutral, V$\,\sim\,$0 (i.e., when excitons cannot bind to holes and become trions). We will show that the exciton-hole scattering enhances the exciton's energy relaxation without rendering it dark.  As a result, bright excitons can get to the light cone faster and emit light before they turn dark by emitting certain types of zone-edge or zone-center phonons \cite{Dery_PRB15}. 

A second often-seen phenomenon, which will be explained by the trion-electron interaction, is a dramatic rather than gradual increase in the PL intensity of negatively-charged bright trions, $X^-_S$ and $X^-_T$. As shown in Fig.~\ref{fig:map}, the dramatic increase in the PL intensity of $X^-_S$ and $X^-_T$ is seen as soon as electrons are added to the ML, followed by decay of the PL intensity when the electron density continues to increase \cite{Chen_NatComm18,Ye_NatComm18,Li_NatComm18,Barbone_NatComm18,Liu_PRL19, Li_NanoLett19}. We will show that the trion-electron scattering renders the bright negative trions dark, and therefore, it shortens the  lifetime of bright trions. The interplay between  lifetime shortening and formation time of bright trions will be used to explain the empirical findings. 

In addition, we will elucidate why the PL from the elusive indirect exciton is stronger when the photoexcitation is circularly polarized ($I^0$ and its phonon replicas $I^0_1$ and $I^0_3$ in Fig.~\ref{fig:map}). Because the photoexcitation in this case depletes electrons from one of the bottommost valleys through dynamical valley polarization \cite{Robert_NatComm21}, the interaction between indirect excitons and electrons from the depleted valley is suppressed. Consequently, exchange scattering that relaxes indirect excitons to dark ones is inhibited, explaining why the total PL intensity of indirect excitons is stronger when the photoexcitation is circularly polarized compared with linearly polarized.

\subsection{Organization of this work}

The organization of this paper is as follows. Section \ref{sec:back} is a comprehensive background of the properties of various excitonic species in ML-WSe$_2$. We add a few new insights that will help us understand the contexts of the experimental and theoretical analyses we present in the two sections that follow. In Sec.~\ref{sec:exp}, we present empirical observations that are central to this work and then present the theory for the exciton-electron and trion-electron interactions in Sec.~\ref{sec:theory}, focusing on exchange scattering processes through which bright excitonic complexes turn dark. Section \ref{sec:results} includes calculated results for the ultrafast relaxation caused by the exciton-electron and trion-electron interactions, and a discussion that clarifies the experimental observations. Finally, we conclude our study and provide outlook. The Appendices include technical details and a compiled list of parameter values used in the simulations.

\section{Background}\label{sec:back}

\begin{figure}
\includegraphics[width=9cm]{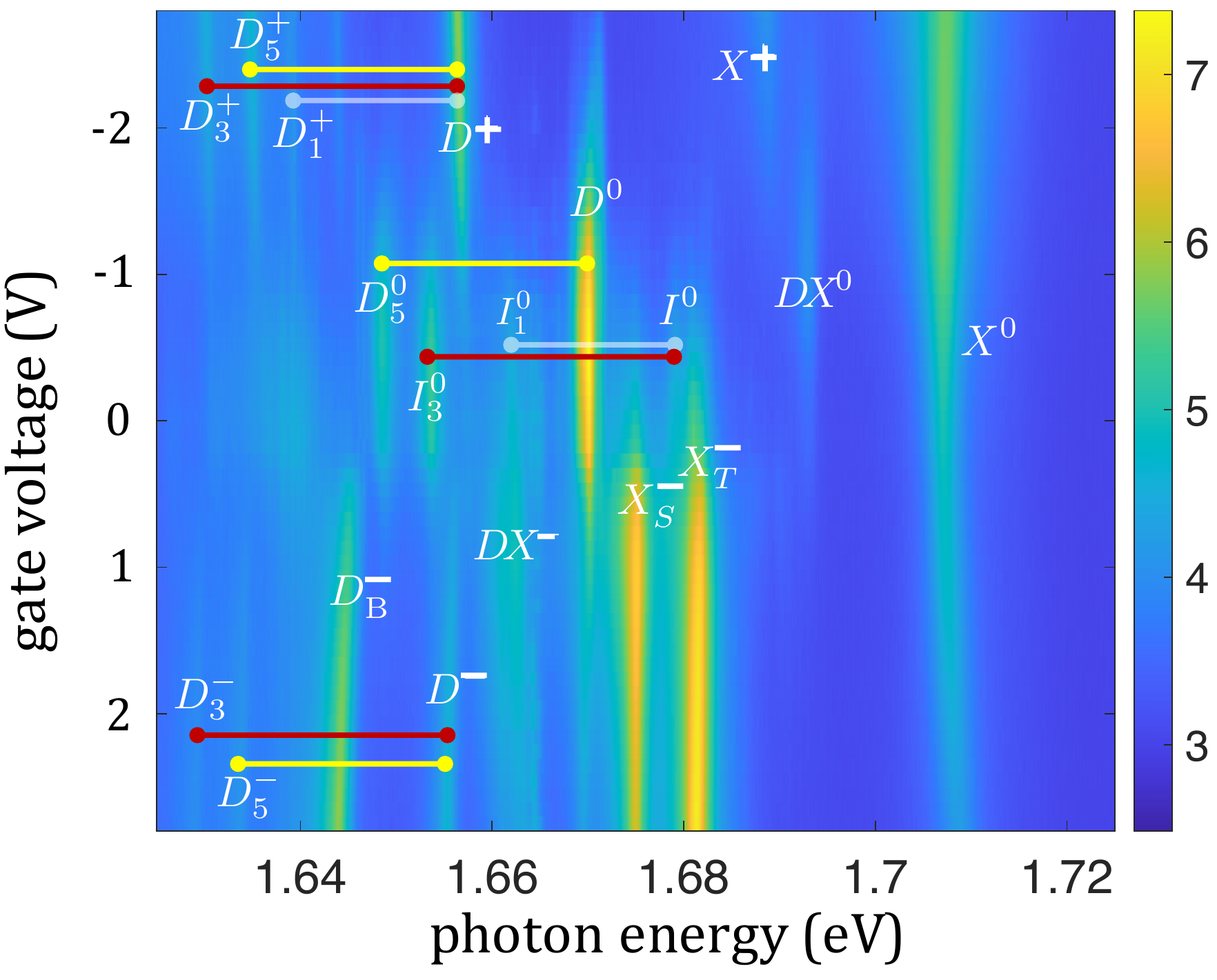}
 \caption{Colormap of the total PL intensity (log scale), following circularly-polarized photoexcitation  of ML-WSe$_2$ that is encapsulated by hexagonal boron-nitride (hBN). The temperature is 4~K, the excitation energy is 1.96 eV, and the laser power is 3.2~$\mu$W. We elaborate on the nature of the marked spectral features throughout Sec.~\ref{sec:back}. Important phonon replica of dark and indirect complexes, marked with $D$ and $I$, are indicated by the horizontal lines, where the corresponding phonon energies are about 26, 21, and 18 meV for the modes $K_3$, $\Gamma_5$, and $K_1$, respectively.  Details of the sample fabrication and experimental setup can be found in  Ref.~\cite{Robert_NatComm21}. } \label{fig:map}
\end{figure}

The PL spectrum of ML-WSe$_2$ is exceptionally rich at low temperatures because of a miscellany of bright and non-bright excitonic complexes, as shown in Fig.~\ref{fig:map}. The PL includes strong emission not only from bright complexes, labeled with $X$ marks, but also from dark and indirect ones ($D$ and $I$ marks). The bright complexes in this material have higher energy, and hence, most of them relax to non-bright complexes after their formation. The signature of dark and indirect species is seen in the PL because their relatively large population compensates their weak optical transition. On the other hand, the signature of bright species is seen in the PL because their strong optical transition compensates their small population.  

\subsection{Bright complexes}\label{sec:x}

\begin{figure}
\includegraphics[width=8cm]{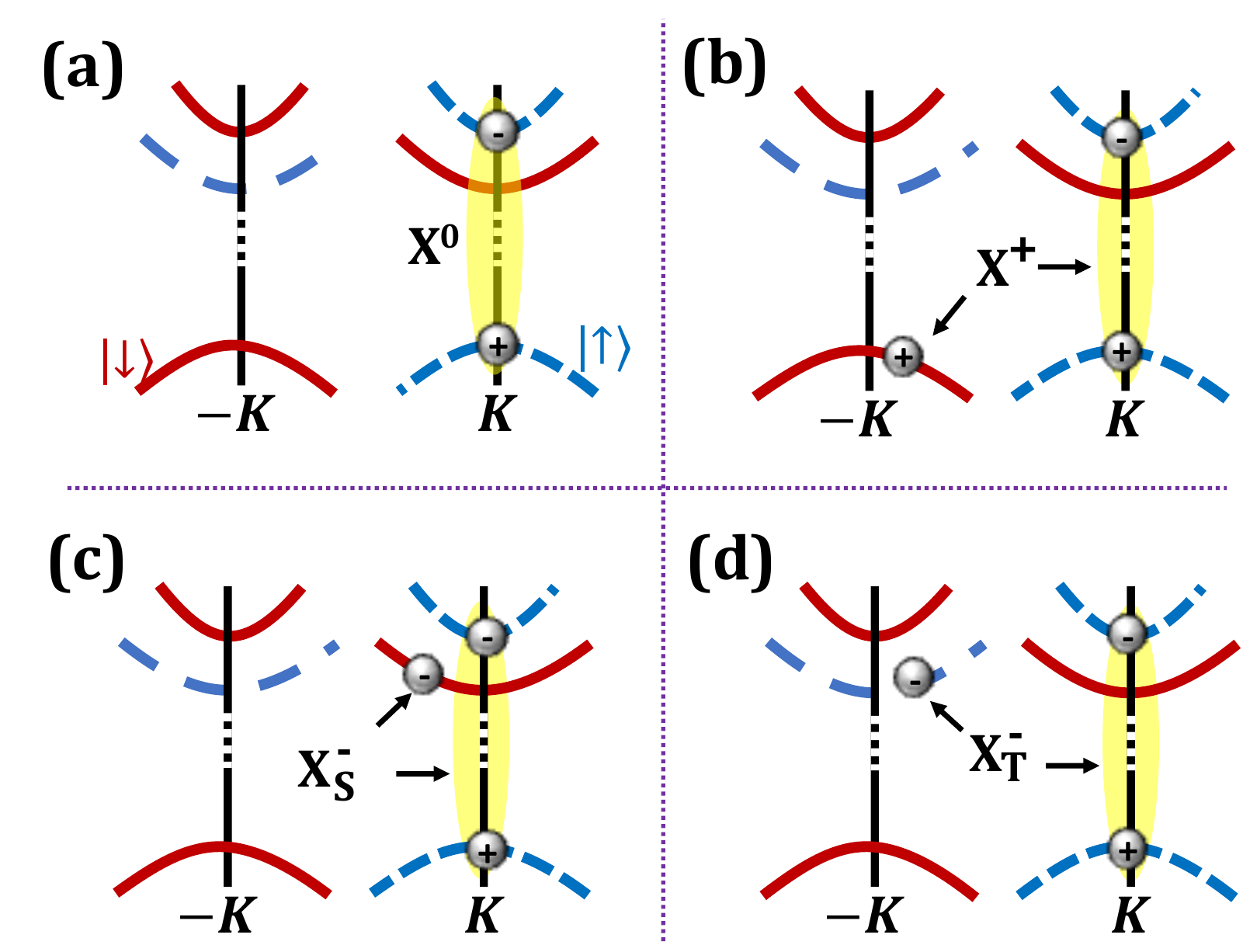}
 \caption{State composition of bright excitonic complexes in ML-WSe$_2$. (a) Neutral exciton $X^0$, (b) positive trion $X^+$, (c) intravalley (singlet) negative trion $X^-_S$, and (d)  intervalley (triplet) negative trion $X^-_T$.} \label{fig:bright}
\end{figure}

Starting with the bright species, Fig.~\ref{fig:bright} shows their state composition in ML-WSe$_2$, where the optically active electron belongs to the top spin-split valley of the CB. The binding between the electron and hole of a bright exciton, shown in Fig.~\ref{fig:bright}(a), is of the order of 160-190~meV in hBN-encapsulated ML-WSe$_2$ \cite{Stier_PRL18,Liu_PRB19,Goryca_NatComm19,Kapuscinski_NatComm21}. Given that the bright exciton optical transition is slightly higher than 1.7~eV in Fig.~\ref{fig:map}, the optical  band gap for free electron-hole pair  transitions in the continuum is $\sim$1.9~eV \cite{Liu_PRB19,Stier_PRL18,Kapuscinski_NatComm21}.

Figures~\ref{fig:bright}(c) and (d) show the negative bright trion complexes, wherein an electron from one of the bottommost valleys binds to the bright exciton. As a result, we can have two bright species depending on the quantum numbers of the two electrons in the complex:  opposite spins but similar valley (intravalley or singlet trion, $X^-_S$), or opposite valleys but similar spin (intervalley or triplet trion, $X^-_T$) \cite{Jones_NatPhys16}. The energy gained by binding of the electron to the bright exciton component is $\sim$35 and $\sim$ 28~meV for the singlet and triplet species, as can be seen from the energies of $X^0$, $X^-_T$ and $X^-_S$ in Fig.~\ref{fig:map}.  The difference in binding energies comes from the dependence of the short-range Coulomb interaction on the valley and spin degrees of freedom \cite{Courtade_PRB17,Hichri_PRB20,Glazov_JCP20}. Positive bright trion complexes, $X^+$, are formed when the exciton binds to a hole from the time-reversed topmost valley of the VB, as shown in Fig.~\ref{fig:bright}(b). The energy gained by this binding is $\sim$20~meV, as can be seen from the energies of $X^0$ and $X^+$ in Fig.~\ref{fig:map}. 

The larger binding energy of negative bright trions is unique to tungsten-based MLs because the two electrons have different masses.  The lighter electron comes from the top spin-split valley of the CB  and the heavier electron comes from the bottom one ($m_{e,t}\,=\,$0.29$m_0$ and $m_{e,b}\,=\,$0.4$m_0$ in ML-WSe$_2$ \cite{Kormanyos_2DMater15}). On the other hand, the masses of the two holes in a positive trion are equal, because both holes come from the topmost time-reversed valleys of the VB ($m_{h}\,=\,$0.36$m_0$ \cite{Kormanyos_2DMater15}). These differences between negative and positive bright trions are consequential since the binding energy of a trion is enhanced/suppressed when the added charge is heavier/lighter than the one with similar charge in the neutral exciton \cite{VanTuan_PRB18}. This difference explains why the measured binding energy of negative bright trions is larger than that of the positive one.

\begin{figure}
\includegraphics[width=8cm]{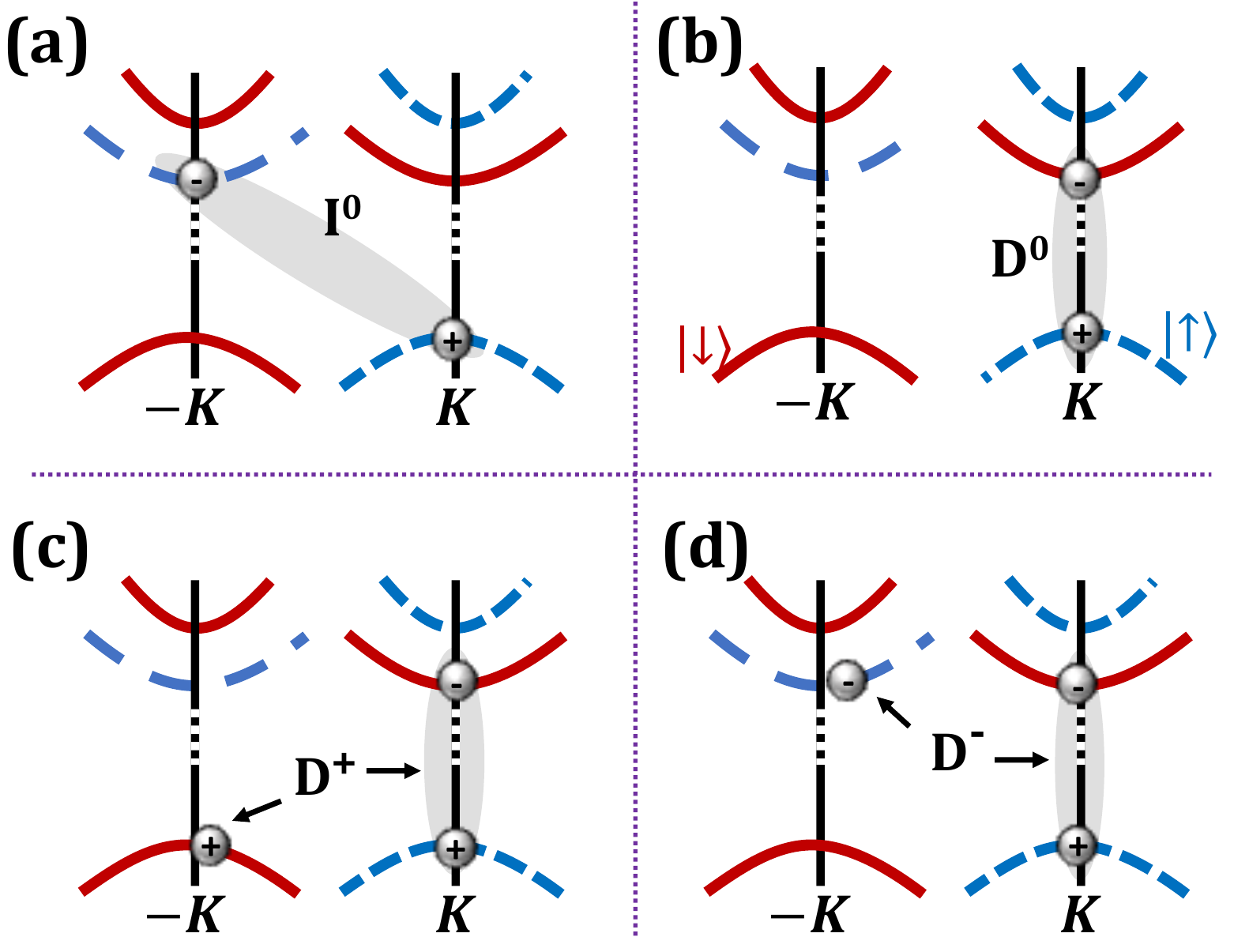}
 \caption{State composition of non-bright excitonic complexes in ML-WSe$_2$, including (a) the indirect exciton, $I_0$, (b) dark exciton, $D^0$, (c) positive dark trion, $D^+$, and (d) negative dark trion, $D^-$.} \label{fig:dark}
\end{figure}


\subsection{Non-bright complexes}\label{sec:xt}

Turning our attention to non-bright species, Fig.~\ref{fig:dark} shows their state composition in ML-WSe$_2$, wherein electrons come from the bottommost CB spin-split valleys. We refer to indirect excitons in this work as ones wherein the electron and hole reside in different valleys, as shown in Fig.~\ref{fig:dark} (a). This nomenclature follows the good-old definition of momentum indirect excitons in multivalley semiconductors such as Si and Ge \cite{Cardona_book}. The reader should not confuse the indirect exciton jargon we use here with other familiar jargons that refer to indirect excitons in real-space where the electron and hole reside in different material layers (known as type-II excitons), or to momentum dark excitons in which a bright exciton is outside the light cone. The other three excitonic complexes in Figs.~\ref{fig:dark}(b)-(d)  are the dark exciton, positive and negative dark trions. 

The indirect exciton, $I^0$ in Fig.~\ref{fig:map}, emerges as a weak signal in the low energy shoulder of the triplet trion (see also Fig.~\ref{fig:I}). Its identification can be verified through a combination of its distinct $g$-factor and polarization analysis of its phonon replicas \cite{He_NatComm20,Liu_PRL20}. Comparing the energies of $D^0$ and $I^0$ in Fig.~\ref{fig:map}, the indirect exciton emerges $\Delta_{id}\,\sim\,$9~meV above the dark one. Given that the electron mass is the same in both species, the smaller binding of the indirect excitons comes only from the electron-hole exchange interaction \cite{Qiu_PRL15,Echeverry_PRB16,Deilmann_PRB17}. This short-range interaction is effective when the spin of the electron in the CB is parallel to that of the missing electron in the VB. While the interaction is repulsive in nature,  it is still much weaker compared with the direct Coulomb attraction between the electron and hole. As a result, the binding energy of the indirect exciton is slightly smaller than that of the dark one. 

The positive dark trion complex, $D^+$, is formed when the dark exciton binds to a hole from the time-reversed valley, as shown in Fig.~\ref{fig:dark}(c). Similarly, the negative dark trion complex, $D^-$, is formed when the dark exciton binds to an electron from the time-reversed valley, as shown in Figs.~\ref{fig:dark}(d). The gained energy from the binding is $\sim$13 meV for the positive dark trion and $\sim$14~meV for the negative one, as seen from the energies of $D^0$, $D^+$ and $D^-$ in Fig.~\ref{fig:map}.  The binding energies are nearly the same because (i) the masses of the two holes (electrons) in the positive (negative) dark trion are the same, and (ii) the electron mass in the bottommost CB valley is only slightly heavier than the hole mass in the topmost VB valley.

In addition to two- and three-body complexes, Fig.~\ref{fig:map} shows emission from neutral and charged biexcitons, denoted by $DX^0$ and $DX^-$. We will not analyze these complexes in this work, and interested readers can refer to Refs.~\cite{Chen_NatComm18,Ye_NatComm18,Li_NatComm18,Barbone_NatComm18}.

.  

\subsection{The spin-splitting energy of the CB ($\Delta_c$)}\label{sec:delc}

The spin-splitting energy between the top and bottom CB valleys, $\Delta_c$, will be shown to be a crucial factor when one tries to analyze the relaxation dynamics of excitonic complexes. First principle calculations provide scattered values for this parameter, ranging from 7~meV up to 40~meV \cite{Kormanyos_2DMater15,Kosmider_PRB13,Echeverry_PRB16}. Unfortunately, this variance is larger than the resolution needed in order to understand the dynamics of excitons after photoexcitation. A recent study by Kapu\'{s}ci\'{n}ski \textit{et al.} reported that $\Delta_{c}\,\simeq\,14\,$~meV, in which the spectrum of bright and dark exciton states was probed near the free electron-hole pair continuum \cite{Kapuscinski_NatComm21}. Larger values  were reported in previous experimental studies \cite{Zhang_NatNano17,Wang_NanoLett17}, in which the spin-splitting energy was only indirectly extracted. Because this parameter is important for the analysis we provide below, we elaborate on this point and estimate the amplitude of $\Delta_c$ using two different considerations; both of which agree well with the reported result in Ref~\cite{Kapuscinski_NatComm21}.

One way to estimate the amplitude of $\Delta_c$ is to use the following analysis of neutral excitons. The optical transition of the bright exciton ($X^0$) in Fig.~\ref{fig:map} is higher than that of the dark one ($D^0$) by $\Delta_{\text{bd}}\,\sim\,$40~meV \cite{Wang_PRL17}. Because the band-gap energies of these transitions differ by $\Delta_c$, the dark-exciton  binding is larger than that of the bright one by $\Delta_{\text{bd}} - \Delta_{c}$. Similarly, the indirect-exciton binding is larger than that of the bright one by $\Delta_{\text{bi}} - \Delta_{c}$, where $\Delta_{\text{bi}}\,\sim\,$31~meV is the measured energy difference between their optical transitions, as shown by Fig.~\ref{fig:I}. The values of $\Delta_{\text{bd}} - \Delta_{c}$ and $\Delta_{\text{bi}} - \Delta_{c}$ are governed by two contributions. The first one comes from exciton mass differences, where bright excitons are lighter due to the smaller effective mass of electrons in the top CB valley. To estimate this effect, one can solve the exciton Schr\"{o}dinger Equation with effective masses taken from first-principles, getting that the bright-exciton binding energy is smaller by $\sim$15-17~meV in hBN-encapsulated ML-WSe$_2$ \cite{VanTuan_PRB18}.  The second contribution is from the electron-hole exchange interaction, relevant for bright and indirect excitons wherein the spin of the CB electron is parallel to that of the missing electron in the VB. Because of the short-range nature of this repulsive interaction, the lowering in binding energy should be similar for the indirect and bright excitons. Therefore, it can be estimated from the energy difference between the dark and indirect exciton optical transitions ($\sim\,9\,$~meV).  These arguments imply that $\Delta_{\text{bi}} - \Delta_{c}$ is mostly governed by the mass difference effect, whereas $\Delta_{\text{bd}} - \Delta_{c}$ has contributions from both mass and electron-hole exchange effects. Putting these pieces together with the aforementioned mass-difference contribution, we estimate that $\Delta_{c} \sim 14-16$~meV.

Another  way to evaluate $\Delta_c$ comes from the brightened emission of the dark trion, denoted by $D_B^-$ in  Fig.~\ref{fig:map}. The brightening is caused by the spin-valley mixing of the two electrons in the dark trion (Figs.~\ref{fig:dark}(d)), where light is generated from recombination of the indirect exciton component \cite{Danovich_SR17,Tu_JPCM19}. To conserve momentum, the left behind electron ends up in the opposite top valley (i.e., intervalley spin-conserving transition). Because of the latter, the $g$-factor of the optical transition ends up being similar to that of bright rather than dark trions. Energy conservation of this physical process means that the spectral line $D_B^-$ emerges below $D^-$, and $\sim$$\Delta_c$  should be their energy difference.  The energy difference is 12~meV according to Fig.~\ref{fig:map}, which agrees quite well with the lower estimate of $\Delta_c$.

\begin{figure*}
    \begin{center}
    \includegraphics[width=15 cm]{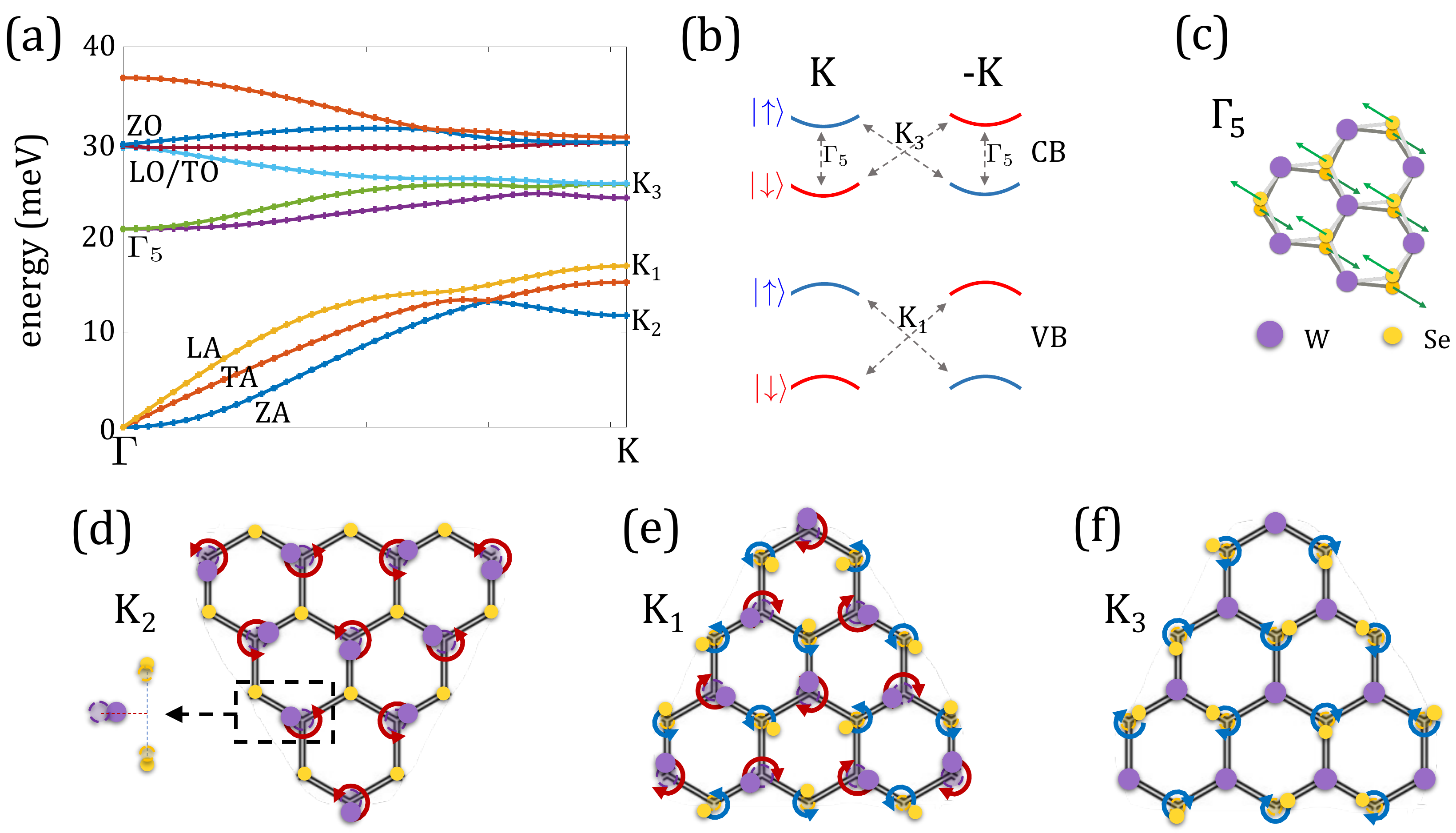}
    \caption{(a) The phonon dispersion of ML-WSe$_2$ along the axis between the high-symmetry $\Gamma$ and $K$ points. Zone-center and pertinent zone-edge phonon modes are indicated. (b) Scheme of low-energy valleys in the CB and VB. Spin-flip intravalley transitions in the CB are mediated by the phonon-mode $\Gamma_5$, while spin-conserving intervalley transitions in the CB (VB) are mediated by the phonon mode $K_3$ ($K_1$). Phonon-induced intervalley spin-flip scattering is relatively weak (the transition matrix element between time-reversed states vanishes). (c) Atomic displacement of the zone-center  phonon mode $\Gamma_5$. (d)-(f) Atomic motions of zone-edge phonon modes that are marked in (a). The curved arrows denote the in-plane circular motion of atoms around their equilibrium positions.} \label{fig:phonon}
    \end{center}
\end{figure*}

We strengthen this estimate by addressing the following question. Would a dark exciton prefer binding to free electron from the top valley to form a singlet trion or to a free electron from the bottommost valley to form a dark trion?  According to Fig.~\ref{fig:map}, the optical transition of $D^0$ emerges $\sim\,$5~meV below $X_S^-$ and $\sim\,$14~meV above $D^-$. That is, the gained binding energy by forming a singlet trion is $\Delta_c - 5$~meV, versus 14~meV by forming a dark negative trion. One can then conclude that the formation of a singlet trion  would be preferable if  $\Delta_c \gtrsim\,$19~meV, and that of a dark negative trion if $\Delta_c \lesssim\,$19~meV.  The fact that dark trions are observed in PL, as shown in Fig.~\ref{fig:map}, reinforces the possibility that $\Delta_c \lesssim\,$19~meV. Namely, dark trions have the lowest energy, and hence their density can be relatively large enough to offset their weak dipole transition, resulting in a noticeable signature in PL experiments.


\subsection{Phonon-assisted optical transitions of indirect and dark excitonic complexes}\label{sec:ph_ass}

When indirect and dark excitonic complexes interact with certain phonon modes, the complex can transfer to a bright (virtual) intermediate state through which light is emitted effectively \cite{Dery_PRB15}. The horizontal lines in Fig.~\ref{fig:map} connect between the bare optical transitions of the dark/indirect complex and its phonon-assisted  replica. 

Figure~\ref{fig:phonon}(a) shows the phonon energy dispersion in ML-WSe$_2$,  calculated using Quantum Espresso \cite{QE}.  According to selection rules \cite{Song_PRL13,Dery_PRB15}, the modes marked by $\Gamma_5$, $K_1$ and $K_3$ in Fig.~\ref{fig:phonon}(a) are those that induce relatively strong transitions between different electronic states in the CB or VB, as shown in Fig.~\ref{fig:phonon}(b). The coupling between opposite spin states in the CB is governed by the long wavelength (intravalley) phonon mode $\Gamma_5$. Its energy is ~$\sim$21~meV in ML-WSe$_2$ and its atomic vibration is shown in Fig.~\ref{fig:phonon}(c), corresponding to out-of-phase and in-plane motion of the two chalcogen atoms in the unit cell. In addition, Fig.~\ref{fig:phonon}(b)  shows that spin-conserving intervalley selection rules are governed by the zone-edge phonon mode $K_3$ for electrons and  $K_1$ for holes. The energies of these phonons in ML-WSe$_2$ are ~$\sim$26~meV and ~$\sim$17~meV, respectively, and their atomic vibrations involve circular motion of the atoms in the unit-cell as shown in Figs.~\ref{fig:phonon}(e) and (f). As shown in Fig.~\ref{fig:map}, the signature of $K_3$ is stronger in PL compared with that of $K_1$. The reason is that the spin-splitting energy in the CB is much smaller, thereby enabling better energy matching between the initial dark complex and intermediate (virtual) bright one ~\cite{He_NatComm20}. 


\subsection{Phonon-assisted darkening of bright excitonic complexes}\label{sec:ph_rel}

When energy relaxation is governed by electron-phonon interactions, spin-conserving relaxation is associated to the gradient of the spin-independent crystal potential, while spin-flip interactions are associated to the gradient of the spin-orbit interaction \cite{Song_PRB12}. As long as the electronic states are not strongly spin-mixed, the spin conserving processes are typically stronger. Recently, He \textit{et al.} analyzed the dark trions’ polarization in ML-WSe$_2$ and showed that spin-conserving intervalley relaxation, mediated by zone-edge phonons ($K_3$), is indeed stronger than spin-flip intravalley relaxation that is mediated by zone-center phonons ($\Gamma_5$) \cite{He_NatComm20,Liu_PRL20}. Yet, the effect of the spin-orbit coupling in these MLs cannot be ignored due to the heavy tungsten atoms.  Below, we identify the energetically permissible relaxation processes through which these phonons are emitted ($K_3$ and $\Gamma_5$), rendering bright excitonic complexes indirect or dark \cite{Dery_PRB15}. 

The bright neutral exciton relaxes to indirect exciton when its electron component interacts with the lattice and a zone-edge phonon is emitted ($K_3$), or it can relaxes to dark exciton through emission of a zone-center phonon ($\Gamma_5$). Both relaxation channels are energetically favorable leading to hot dark or indirect excitons because $\Delta_{\text{bd}} > E_5$ and $\Delta_{\text{bi}} > E_3$, where $E_5\,\sim\,21$~meV and $E_3\,\sim\,26$~meV are the respective phonon energies. As before, $\Delta_{\text{bd}}\,\sim\,40$~meV ($\Delta_{\text{bi}}\,\sim\,31$~meV) corresponds to the energy difference between the optical transitions of the bright and dark (indirect) excitons \cite{footnote1}.  

Switching to permissible trion relaxation processes, the bright positive trion becomes dark when its  lone electron interacts with the lattice and a phonon is emitted (either $K_3$ or $\Gamma_5$). As can be seen from the spectral lines of $X^+$ and $D^+$ in Fig.~\ref{fig:map}, the energy difference between the bright and dark positive trions is $\sim\,$33~meV. $E_3$ and $E_5$ are both smaller than this difference, implying that the phonon-induced relaxation of a positive bright trion to dark one is effective. Next, the bright triplet  trion relaxes to dark trion when the optically active electron of the triplet interacts with the lattice and a zone-center phonon is emitted ($\Gamma_5$ ). As can be seen from Fig.~\ref{fig:map}, the energy difference between the spectral lines of $X^-_T$ and $D^-$ is $\sim\,$26~meV, which is larger than $E_5$.  Thus, the relaxation of bright triplet trions to dark ones is also energetically favorable. 

The one exception comes from the bright singlet negative trion. According to Fig.~\ref{fig:bright}(c) and the selection rules in Fig.~\ref{fig:phonon}(b), the singlet complex should turn dark when its optically active electron interacts with the lattice and a zone-edge phonon $K_3$ is emitted. Yet, unless the bright trions are hot, this relaxation is not energetically permissible.  The energy difference between the spectral lines of $X^-_S$ and $D^-$ in Fig.~\ref{fig:map} is $\sim\,$19~meV, which is smaller than $E_3$. As a result, the phonon-induced relaxation of singlet negative trions is hampered at low temperatures. The relaxation can eventually be established by emission of zone-edge phonon modes with energy smaller than $\sim\,$19~meV, such as the mode $K_2$ in Figs.~\ref{fig:phonon}(a) and (d). However, this process is relatively slow because it is not allowed by (lowest-order) selection rules. We will see in the next section that this unique behavior of the singlet trion is supported by time-resolved PL experiments.

\begin{figure*}
\includegraphics[width=17cm]{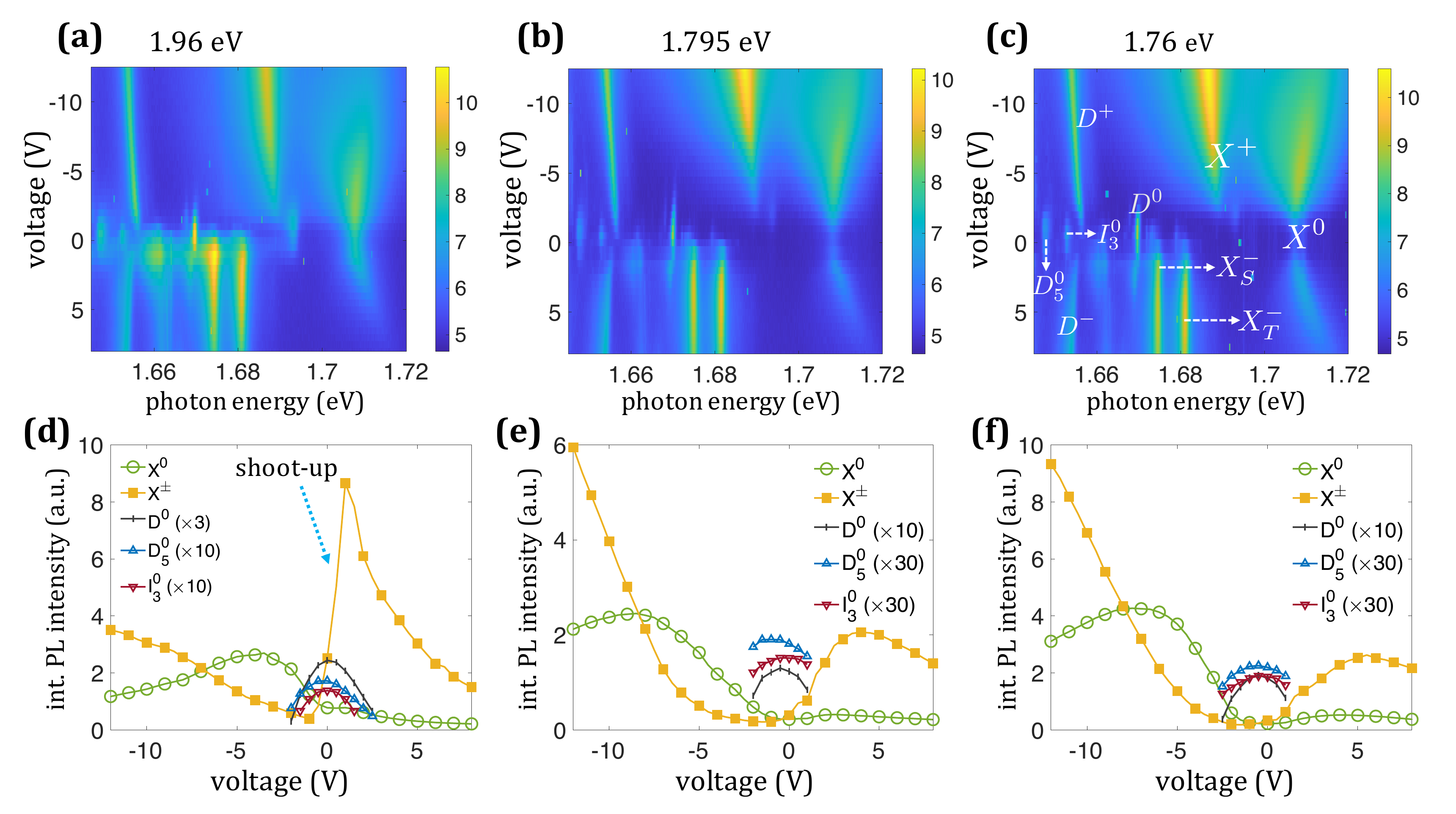}
 \caption{(a)-(c) Colormaps of the total PL intensities (log scale) in a second ML-WSe$_2$ charge-tunable device, following linearly-polarized photoexcitation. The temperature is 4~K and the laser power is 10~$\mu$W. The laser photon energies are indicated in each case, where $\hbar\omega\,=\,$1.96~eV in (a) corresponds to excitation of free electron-hole pairs in the continuum. The other two laser energies in (b) and (c) fall in the spectral range between the ground state of the bright exciton and excited-state of bright trions (i.e., between 1$s$ and 2$s^{\pm}$) . (d)-(f) The integrated PL intensities of various peaks, calculated from the respective PL spectra in (a)-(c). Results are shown for the bright exciton (open circles), bright trions (filled squares), dark exciton, its $\Gamma_5$ phonon replica (triangles), and the $K_3$ phonon replica of the indirect exciton (flipped triangles). } \label{fig:various}
\end{figure*}

 \begin{figure}[b!]
\includegraphics[width=8cm]{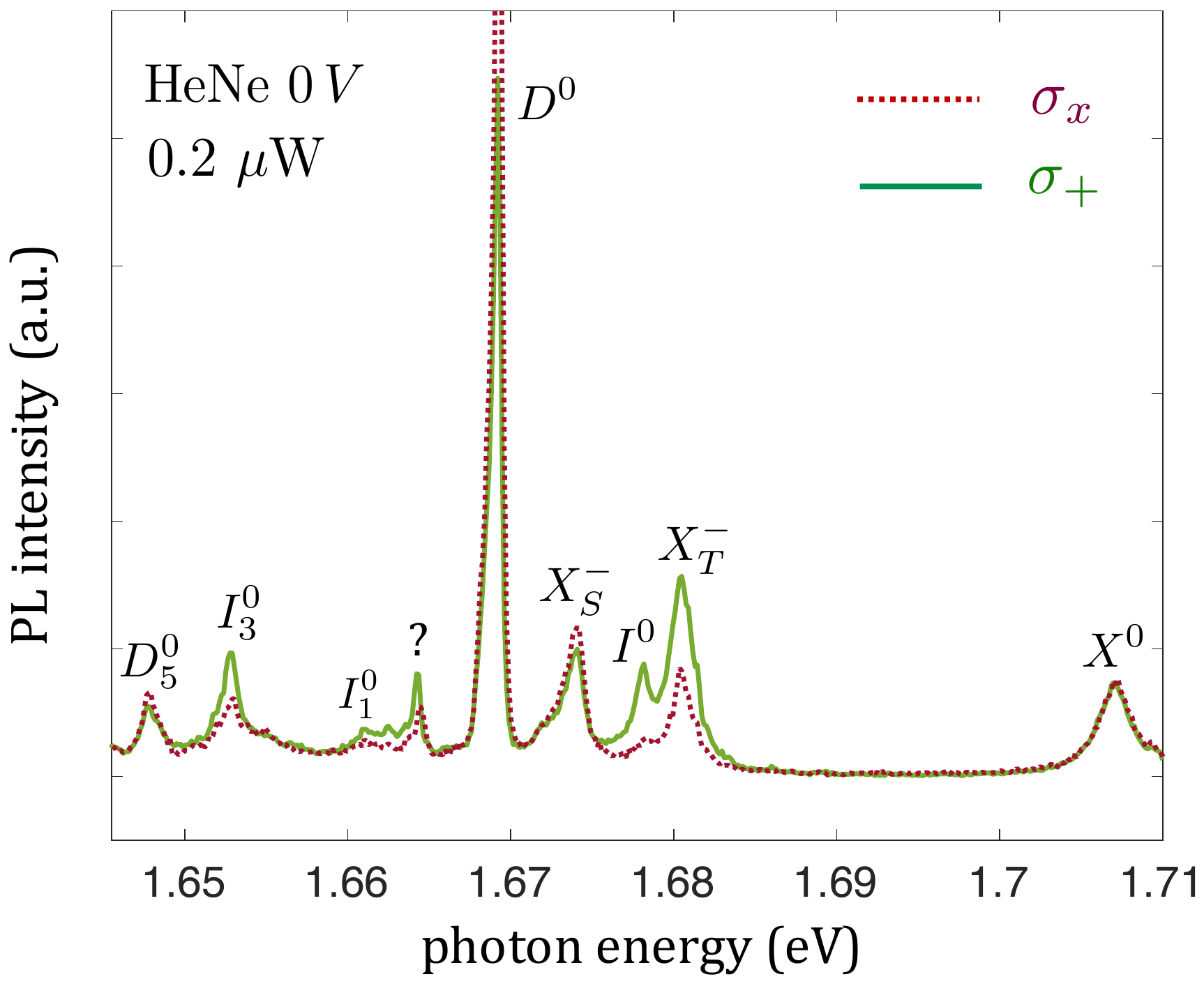}
 \caption{PL intensity of ML-WSe$_2$ at zero gate voltage. The temperature is 4~K and the excitation energy is 1.96 eV. Solid (dotted) lines correspond to the PL when the laser is circularly- (linearly-) polarized. The bare optical transition of the indirect exciton ($I^0$) is  best resolved from the trion triplet peak ($X^-_T$) at relatively low powers (0.2~$\mu$W in this case).
 } \label{fig:I}
\end{figure}
 
\section{PL spectra as a function of electrostatic doping}\label{sec:exp}

The previous discussion highlighted the exceptionally rich PL spectrum of high-quality ML-WSe$_2$ at low temperatures. Hereafter, we will be focusing on a series of measurements through which the interaction between excitonic complexes and electrons or holes is unraveled. We present polarization dependent micro-PL experiments at 4~K, and unless otherwise stated, the excitation source is the 632.8~nm line of a HeNe laser (1.96 eV; i.e. excitation above the free-carrier band gap energy), focused to a spot size smaller than 1~$\mu$m diameter. The neutrality point of the device is reached when the gate voltage is $\sim$0.5~V, measured with complementary differential reflectivity experiments (not shown). A change of 1~Volt in the devices we work with corresponds to a change of ~10$^{11}$~cm$^{-2}$ in charge density. Results of time-resolved PL experiments are carried using a ps-pulsed laser (TiSa) at a wavelength of 695~nm (1.78 eV) whose full-width at half maximum is $\sim$2~ps, the repetition rate is 80~MHz,  and average power is 5~$\mu$W. The PL signal is detected by a Hamamatsu streak camera with a time resolution of ~2-3 ps. Further details can be found in Ref.~\cite{Robert_NatComm21}. 


The first observation we wish to focus on is that \textit{the PL intensity of the neutral bright exciton, $X^0$, is strongest when the ML is hole-doped rather than charge neutral}, as shown in Fig.~\ref{fig:map}.  A similar behavior is inspected when the ML is excited at other laser energies, as shown in Fig.~\ref{fig:various}. In all shown cases, the PL intensity of $X^0$ is much stronger when the ML is hole doped compared with the  neutral and electron-doped regimes (the PL colormaps are in log scale). This behavior is further evidenced from the integrated PL of the neutral exciton, as shown by the solid lines with empty circle markers in Figs.~\ref{fig:various}(d)-(f). Note that the stronger PL intensity in the hole-doped regime was observed by other groups but not commented \cite{Chen_NatComm18,Ye_NatComm18,Li_NatComm18,Barbone_NatComm18,Liu_PRL19, Li_NanoLett19}. Contrary to the non-trivial dependence of the PL intensity of $X^0$ on charge density, Figs.~\ref{fig:various}(d)-(f) show that the PL intensities of dark and indirect excitons behave as one would expect in the sense that they are strongest when the ML is charge neutral.

Another unique behavior, which is readily visible in Figs.~\ref{fig:map} and \ref{fig:various}, is a \textit{fast rather than gradual increase in the PL intensity of bright negative trions, $X^-_S$ and $X^-_T$}.  Furthermore, Fig.~\ref{fig:various}(d) shows that the integrated PL shoots up dramatically as soon as electrons are added to the ML when electron-hole free pairs are excited in the continuum (i.e., excitation with HeNe laser at $\hbar\omega\,=\,$1.96~eV). The shoot-up is governed by emission from bright negative trions, and it decays when the charge density is a few times 10$^{11}$~cm$^{-2}$. Contrary to this nontrivial behavior,  the PL intensity of positive trions increases gradually without a visible decay even at relatively large hole densities. This behavior is independent of the laser excitation energy, as shown in  Figs.~\ref{fig:various}(d)-(f).

The last observation we make is more subtle and deals with the polarization dependence of the PL intensity of  indirect excitons. Figure~\ref{fig:I} shows the total PL intensity emitted by the ML following either circular excitation (solid line) or linear excitation (dotted line).  As expected, the PL intensity of the bright exciton, $X^0$, is similar for linear and circular excitation. On the other hand, the emission of the triplet trion, $X^-_T$, is twice more intense following circular excitation as a consequence of  dynamical valley polarization \cite{Robert_NatComm21}. Interestingly, we observe a similar behavior for the PL intensity of the indirect exciton $I^0$, as shown Fig.~\ref{fig:I}. Namely, \textit{the peaks $I^0$ and its dominant phonon replica  $I^0_3$ are more salient under circularly polarized excitation}. 

\section{Theory}\label{sec:theory}
We will show that all of the observations discussed so far can be understood with an overarching argument: exchange scattering between excitonic complexes and free charge particles.  This section sets forth the theory for scattering between electrons and exciton complexes, largely following the work of Ramon \textit{et al.} for semiconductor quantum wells \cite{Ramon_PRB03}, and of Shahnazaryan \textit{et al.}  for ML-TMDs \cite{Shahnazaryan_PRB17}. The important addition we emphasize in this work deals with the character change of excitons  following exchange scattering with electrons (e.g., bright exciton relaxation to dark or indirect exciton following scattering with a free electron). Later, we will analyze the trion-electron exchange scattering to better understand how bright negative trions in tungsten based ML-TMDs turn dark following interaction with free electrons. 

The reader should not confuse between the two exchange terms we use in this work. The short-range and repulsive electron-hole exchange interaction we have discussed before has to do with the slight mixing of CB and VB states. It gives rise to slightly smaller exciton binding energy. The exchange scattering we will study below has to do with Coulomb scattering between excitonic complexes and free electrons (or holes), after which the particle composition of the complex changes.

\subsection{Exciton-electron scattering}\label{sec:xe}

A convenient way to describe the scattering process is by writing the combined exciton and electron  states as $| \alpha, \mathbf{k}_x ; j, \mathbf{k}_e \rangle$. Here, $\alpha = \{X^0,\,D^0,\,I^0 \}$ is the exciton species and $\mathbf{k}_x$ is its center-of-mass wavevector. The index $j=\{t,b\}$ indicates whether the free electron belongs to the top or bottom valleys of the CB, and $\mathbf{k}_e$ is its wavevector. The scattering matrix element has contributions from Coulomb interactions of the free electron with the hole and electron components of the exciton. Using Fermi's golden rule, the rate at which an exciton of species $\alpha$ with center-of-mass wavevector $\mathbf{k}_x$ relaxes  to species $\beta$ reads,
\begin{widetext}
\begin{eqnarray} \label{eq:rate_3}
R(\mathbf{k}_x; \alpha \!\rightarrow\! \beta  )  =  \frac{2\pi}{\hbar } \sum_{\mathbf{q},\mathbf{k}_e}  && \left| \left\langle  \beta  , \mathbf{k}_x +\mathbf{q} \,\,;\,\, j, \mathbf{k}_e-\mathbf{q} \left|  V_{eh} + V_{ee} \right|   \alpha, \mathbf{k}_x\,;\,b , \mathbf{k}_e  \right\rangle   \right|^2 \!\! f_{b,\mathbf{k}_e} ( 1 - f_{j,\mathbf{k}_e-\mathbf{q}} ) \nonumber \\ && \delta \left ( \varepsilon_{\beta,\mathbf{k}_x + \mathbf{q}}  + (1-\delta_{j,b})\Delta_c + \varepsilon_{j,\mathbf{k}_e -  \mathbf{q}} - \Delta_{\alpha \beta} - \varepsilon_{\alpha,\mathbf{k}_x} -\varepsilon_{b,\mathbf{k}_e } \right)\! ,
\end{eqnarray}
\end{widetext}
where $\Delta_{\alpha\beta} $ is the difference between the optical transition energies of the two exciton species. The free electron in the initial state is assumed to be thermal, hence residing in the bottom spin-split valley of the CB, where its Fermi distribution and energy are $f_{b,\mathbf{k}_e}$ and $\varepsilon_{b,\mathbf{k}_e }$, respectively.  $f_{j,\mathbf{k}_e-\mathbf{q}}$ and $(1-\delta_{j,b})\Delta_c + \varepsilon_{j,\mathbf{k}_e -  \mathbf{q}}$ are the respective parameters of the free electron in the final state. The energy of the electron in the final state takes into account the possibility that a bright exciton can turn dark or indirect (e.g., $\alpha=X^0$ and $\beta=D^0$~or~$I^0$), thereby  leaving a free electron in the top valley after scattering ($j=t$).

The derivation of the matrix element in Eq.~(\ref{eq:rate_3}) is carried by using second quantization and invoking translation symmetry to integrate out the plane waves in the initial and final electron states and center-of-mass motion of the excitons \cite{Ramon_PRB03}. We then get that 
\begin{widetext}
\begin{eqnarray} \label{eq:Vee_eh}
\left\langle  \beta, \mathbf{k}_x \!+\! \mathbf{q} \,\,;\,\, j, \mathbf{k}_e\!-\!\mathbf{q} \,\left|\,  V_{eh} + V_{ee} \, \right| \,  \alpha, \mathbf{k}_x\,;\,j' , \mathbf{k}_e  \right\rangle_{\text{exc}}  &=& \frac{1}{A}\sum_{\mathbf{k}}  V_{\mathbf{k}+\mathbf{k}_e-\mathbf{q}}  \,  \phi_{\eta_{\beta}\mathbf{k}_x - (1-\eta_{\beta})\mathbf{q} + \mathbf{k}}  \left[ \phi^{\ast}_{\eta_{\alpha}\mathbf{k}_x + \mathbf{q} - \mathbf{k}_e } - \phi^{\ast}_{ \eta_{\alpha}\mathbf{k}_x + \mathbf{k} } \right]\,,\nonumber \\
\left\langle  \beta, \mathbf{k}_x \!+\! \mathbf{q} \,\,;\,\, j, \mathbf{k}_e\!-\!\mathbf{q} \, \left| \, V_{eh} + V_{ee} \, \right| \,  \alpha, \mathbf{k}_x\,;\,j' , \mathbf{k}_e  \right\rangle_{\text{dir}}  &=&\frac{\delta_{\alpha,\beta}}{A}  \sum_{\mathbf{k}} V_{\mathbf{q}}  \,  \phi^{\ast}_{\eta_{\alpha}\mathbf{k}_x + \mathbf{k}}  \left[ \phi_{ \eta_{\beta}\mathbf{k}_x + \eta_{\beta}\mathbf{q} + \mathbf{k} } - \phi_{ \eta_{\beta}\mathbf{k}_x - (1-\eta_{\beta})\mathbf{q} + \mathbf{k} }\right]. 
\end{eqnarray}
\end{widetext}
The monolayer area is denoted by $A$. The breakup to exchange and direct parts in the first and second lines follows from the two configurations in Fig.~\ref{fig:scheme}(b).  $V_{\mathbf{q}}$ is the statically screened Coulomb potential [Eq.~(\ref{eq:Vq}) in Appendix~\ref{app:details}]. $\phi_{\mathbf{k}}= \mathcal{F}\left[ \varphi_x(r) \right]$ is the Fourier transform of the exciton wavefunction for the relative motion of the electron and hole, where $\mathbf{r}=\mathbf{r}_e-\mathbf{r}_h$. $\eta_x = m_e/(m_e+m_h)$ is a mass ratio, where $m_h$ is the hole effective mass in the top  valley edge of the VB, and $m_e$ denotes the electron mass. The latter is the effective mass at the edge of the top (bottom) spin-split CB valley if the exciton is bright (non-bright). In the simplest approximation, one can use  $\phi_{\mathbf{k}}= \sqrt{8\pi} a_x/[1 +k^2a_x^2]^{3/2}$ and treat the exciton Bohr radius, $a_x$, as a fitting parameter. The exciton species can be identified from the mass ratio in the wavevector argument ($\eta_{\alpha}$ or  $\eta_{\beta}$).


To compare the direct and exchange terms in Eq.~(\ref{eq:Vee_eh}), we first note that the direct scattering is only applicable when the exciton does not change its character ($\delta_{\alpha,\beta}=0$ in  the second line when $\alpha \neq \beta$). This property is clear given that direct scattering refers to cases in which the complex does not change its particle composition.  Therefore, we consider first  the case that $\alpha=\beta$ in order to compare the exchange and direct terms on equal grounds. This case is relevant in several scenarios when dealing with ML-WSe$_2$. For example, when an exciton scatters with free holes, wherein the hole of the exciton and the free ones are all from the same topmost valley of the VB. The exciton species remains the same after scattering (bright, indirect or dark), regardless of whether the holes are switched or not. Another example is when a dark (indirect) exciton scatters with free electrons from the same bottommost valley of its own electron, thereby leaving the exciton dark (indirect) after scattering. 

A few properties can help us understand why exciton-electron exchange scattering is an effective mechanism, whereas the direct scattering is weaker. On the one hand, the charge neutrality of the exciton is manifested by an offset between electron-electron and electron-hole interactions at long-wavelengths. In exchange scattering, this effect can be seen from the limit $\mathbf{k} \rightarrow \mathbf{q}-\mathbf{k}_e$ in the first line of Eq.~(\ref{eq:Vee_eh}). While the Coulomb potential is strongest in this limit, the terms in  square-brackets cancel each other. The same holds for direct scattering in the second line of Eq.~(\ref{eq:Vee_eh}) when $\mathbf{q}  \rightarrow 0$.  However, a key difference between exchange and direct scattering is that the charge neutrality of the exciton plays a larger role in direct scattering, resulting in an overall strong cancellation between the electron-electron and electron-hole interactions. In other words, the scattering is weaker if the exciton keeps the same electron-hole pair before and after scattering.

To see how exchange scattering is dominant, we examine the matrix element in the first line of Eq.~(\ref{eq:Vee_eh}), noticing that the first term inside the square brackets, $\phi^{\ast}_{\eta_{\beta}\mathbf{k}_x + \mathbf{q} - \mathbf{k}_e }$, can be moved outside of the sum. This term comes from the electron-hole interaction following exchange of the electrons, yielding $\mathbf{k}_e$ instead of $\mathbf{k}$ in the wavevector argument. Consider first the case that $\phi^{\ast}_{\eta_{\beta}\mathbf{k}_x + \mathbf{q} - \mathbf{k}_e }$ is relatively large, applicable in the limit that  $ \eta_{\beta}\mathbf{k}_x + \mathbf{q} - \mathbf{k}_e \ll a_x^{-1}$. After summation over $\mathbf{k}$, the integrated contribution from the second term in square-brackets (electron-electron) cannot offset the contribution from this term (electron-hole). Conversely, the integrated contribution from the second term is larger when $\phi^{\ast}_{\eta_{\beta}\mathbf{k}_x + \mathbf{q} - \mathbf{k}_e }$ is relatively  small, applicable in the limit that $\eta_{\beta}\mathbf{k}_x + \mathbf{q} - \mathbf{k}_e \gg a_x^{-1}$. The offset between the electron-hole and electron-electron interactions is ineffective in these two limits.

Turning to the direct term in the second line of Eq.~(\ref{eq:Vee_eh}), we notice that both terms inside the square brackets include $\mathbf{k}$ in their wavevector arguments. The offset between their integrated  contributions after summation over $\mathbf{k}$ is far more effective. If we focus on cold excitons with similar electron and hole masses ($\mathbf{k}_x \rightarrow  0$ and $\eta \sim 1/2$), the direct scattering suppression is nearly complete due to very effective cancellation between the electron-hole and electron-electron interactions (e.g., the contributions from $\pm \mathbf{k}$ exactly cancel each when $\eta = 1/2$ and $\mathbf{k}_x =  0$). More on the comparison between direct and exchange scattering can be found in Ref.~\cite{Ramon_PRB03}. All in all,  whereas the effective offset between electron-hole and electron-electron interactions gives rise to small contribution from direct scattering, the exchange scattering is effective because the hole can bind to either of the two electrons post scattering.

Finally, we mention that the exchange-scattering rate $R(\mathbf{k}_x; \alpha \!\rightarrow\! \beta  )$ for $\alpha \neq \beta$ and $\alpha = \beta$ is comparable as long as the binding energies of the excitons before and after scattering are comparable (i.e., when $\Delta_{\alpha\beta}$ is small compared with the binding energies). We will analyze these cases more carefully in the next section. Before doing so, we turn our attention to trion-electron scattering.


\subsection{Trion-electron scattering}\label{sec:te}
When the electron density continues to increase, excitons bind to free electrons effectively, and trions take the place of neutral excitons (see Figs.~\ref{fig:map} and \ref{fig:various}). The trion-electron scattering includes contributions from the interaction of the free electron with the hole, $V_{eh}$, and with both electrons of the trion, $V_{ee} = V_{ee_1} + V_{ee_2}$. As such, there is no effective cancelation between $V_{eh}$ and $V_{ee}$, and the direct scattering term is no longer negligible when the trion species is kept after scattering. Yet, we will mostly focus on exchange scattering through which we will be able to understand the experimental results.  

The valley and spin degrees of freedom in tungsten-based ML-TMDs are such that a bright trion (singlet or triplet) can become dark when it interacts with a free electron. The darkening happens if the free electron is from a (bottommost) valley that is not occupied by the electrons of the bright trion. The final state includes the dark trion and a free electron in the top valley. Focusing first on the transition from singlet to dark trions, the exchange scattering rate $R(\mathbf{k}_T; X_S^- \!\rightarrow\! D^-  )$ can be written in a similar fashion to that of the exciton-electron rate in Eq.~(\ref{eq:rate_3}). The needed changes are to replace the translational wavevector of the exciton with that of the trion ($\mathbf{k}_x$ with $\mathbf{k}_T$), and to assign $\alpha=X^-_S$, $\beta=D^-$, and $j=t$. The matrix element of this exchange scattering case reads 
\begin{widetext}
\begin{eqnarray} \label{eq:SD}
\!\!\!\!\!\!&\!\!\!\!\!\!&\!\!\!\!\!\! \!\!\! \left\langle  D^-, \mathbf{k}_T \!+\! \mathbf{q} \,\,;\,\, t, \mathbf{k}_e\!-\!\mathbf{q} \,\left|\,  V_{eh} + V_{ee} \, \right| \,  X^-_S, \mathbf{k}_T\,;\,b , \mathbf{k}_e  \right\rangle   = \frac{1}{A}\sum_{\mathbf{p},\mathbf{k}}  V_{\mathbf{q}_e - \mathbf{p}}  \, \Big\{   \psi_S^\ast(\alpha_t  \mathbf{k}_T +  \mathbf{p} \,,\,\alpha_b  \mathbf{k}_T +  \mathbf{k}) \psi_D(\mathbf{q}_T +  \mathbf{k} \,,\,\mathbf{p} -  \mathbf{q}_D) \nonumber \\  && \qquad \qquad \qquad  + \,\,\,\,\psi_S^\ast(\alpha_t  \mathbf{k}_T +  \mathbf{q}_e \,,\,\alpha_b  \mathbf{k}_T +  \mathbf{k}) \left[ \psi_D(\mathbf{q}_T +  \mathbf{k}  + \mathbf{q}_e - \mathbf{p}     \,,\,\mathbf{p} -  \mathbf{q}_D) - \psi_D(\mathbf{q}_T +  \mathbf{k} \,,\,\mathbf{p} -  \mathbf{q}_D)  \right] \Big\}\,. 
\end{eqnarray}
\end{widetext}
where $V_{ee}=V_{ee_1} + V_{ee_2}$, $\mathbf{q}_e = \mathbf{q} - \mathbf{k}_e$, $\mathbf{q}_T = \alpha_D( \mathbf{q} + \mathbf{k}_T)$, and $\mathbf{q}_D =  \mathbf{q} -  \mathbf{q}_T$. The mass ratio parameters are $\alpha_t = m_t /M_S$, $\alpha_b = m_b /M_S$, and $\alpha_D = m_b /M_D$, where $M_S=m_b+m_t+m_h$ and $M_D=2m_b+m_h$ are the masses of the singlet and dark trions, respectively. $\psi_{S/D}(\mathbf{p},\mathbf{k})$ is Fourier transform of the singlet/dark trion wavefunction with respect to the relative motion of the three particles in the complex. In the simplest approximation, one can use $\psi_{i}(\mathbf{p},\mathbf{k}) = 8\pi a_ib_i / \{  [1 +p^2a_i^2]^{3/2} \cdot  [1 +k^2b_i^2]^{3/2} \}$, and treat $a_i$ and $b_i$ as fitting parameters to denote the trion radii (effective distances of the hole from the two electrons). The matrix element for exchange scattering from triplet to dark trions is similar but with $\psi_S(\mathbf{k},\mathbf{p}) \rightarrow \psi_T(\mathbf{k},\mathbf{p})$ and with switched arguments in the dark trion wavefunction: $\psi_D(\mathbf{p},\mathbf{k}) \rightarrow \psi_D(\mathbf{k},\mathbf{p})$.
\section{Results, discussion and comparison with experiment }\label{sec:results}
Putting the pieces together, we can see that electron-doped ML-WSe$_2$ is unique in that its CB spin-valley configuration enables relaxation of bright excitonic complexes to dark ones through exchange scattering. The relaxation rate depends on the density of free electrons, which one can control through electrostatic doping. Importantly, one should not confuse exchange scattering with regular spin relaxation, caused when an electron or hole of an excitonic complexes  experiences a spin flip. In exchange scattering, the spin configuration changes when the electron is replaced by a new electron with opposite spin. Another important consequence of the CB spin-valley configuration in ML-WSe$_2$ is that the exchange-driven darkening effect  is absent in hole-doped MLs. The reason is that the lone electron of positive trions or excitons is kept in the top valley when free holes exchange with the hole(s) of the exciton (trion). The only way for positive or neutral bright complexes in hole-doped MLs to relax to the lowest-energy dark states is through interaction of their electron with the lattice ($\Gamma_5$ or $K_3$), after which the electron is transferred to the bottommost valley of the CB. 

\begin{figure*}
\includegraphics[width=17cm]{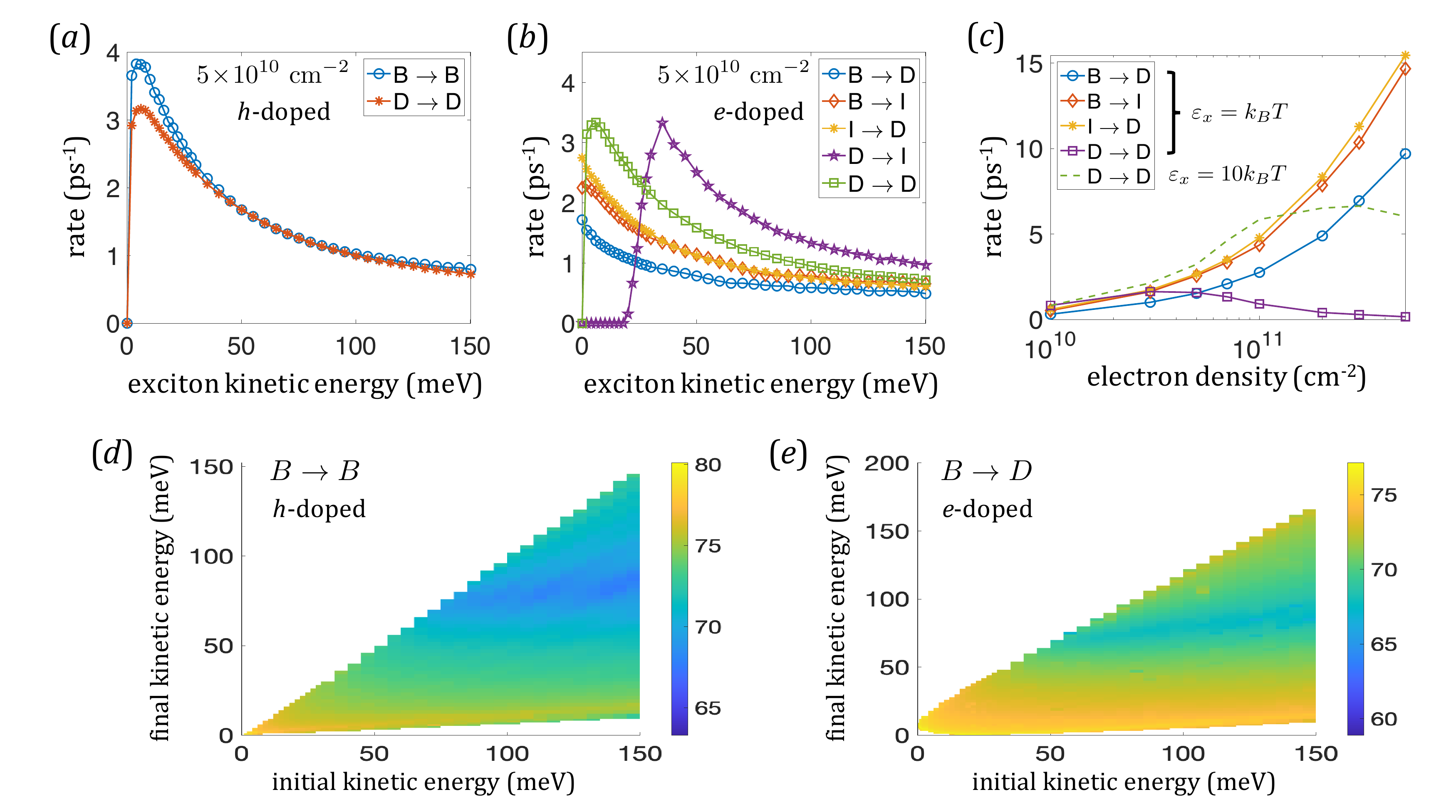}
 \caption{Calculated transition rates between various exciton species in ML-WSe$_2$ at 4$\,$K when the free-charge density is  5$\times$10$^{10}$~cm$^{-2}$. (a)/(b) The rates as a function of the exciton kinetic energy in the initial state for hole/electron doped MLs. Transition rates between indirect excitons ($I \rightarrow I$)  are not shown in these plots but they are similar to the those between dark excitons ($D \rightarrow D$). (c) The rates as a function of electron density for thermal excitons. (d) and (e) Log-scale colormaps of the kernel functions for the transitions $B \rightarrow B$ and $B \rightarrow D$ in hole-doped and electron-doped MLs, respectively. The units of the kernel function are $( J \cdot s)^{-1}$. Yellow/blue colors mean exponentially larger/smaller scattering probability between certain initial ($x$-axis) and final ($y$-axis) kinetic energies.} \label{fig:transitions}
\end{figure*}

\subsection{Excitons}\label{sec:disc_x}

We first present calculations of transition rates between various neutral exciton species in ML-WSe$_2$. Figure~\ref{fig:transitions} shows the results when the charge density in the ML is 5$\times$10$^{10}$~cm$^{-2}$. Transitions from indirect or dark excitons to bright ones (not shown) are ineffective in electron-doped MLs if the density of hot electrons in the top CB valleys is small compared with the density of thermal electrons in the bottommost valleys.

A few immediate conclusions can be drawn from inspection of the results in Fig.~\ref{fig:transitions}.  The first one is that excitons have exceptionally fast relaxation, possibly rendering exchange scattering more efficient in the initial cooling of hot excitons compared with phonon emission processes \cite{Yang_PRB20}. The second remark is that the scattering is suppressed for thermal dark excitons when the charge density increases, as can be seen from the line with open square symbols in Fig.~\ref{fig:transitions}(c). The reason is that Pauli-blocking of electrons (or holes) near the Fermi surface impedes efficient energy and momentum exchange with cold excitons in their ground state.  In cases where the exciton can relax to a different species, the phase-space restriction is alleviated and the excitons cool down very effectively by giving their excess energy to the electrons (or holes). Lastly, whereas  $\Delta_{\text{id}}\,\sim\,9$~meV,  the transition  $D \rightarrow I$ is nonzero when the kinetic energy of the dark exciton is larger than $\sim$20~meV. The excitons have to be that hot to facilitate this transition since thermal electrons with which they scatter have negligible energy and momentum at low densities. The only way to conserve both energy and momentum in this scattering is if the exciton has large enough center-of-mass wavevector, as one can verify from the second line of Eq.~(\ref{eq:rate_3}). 


\subsubsection{Bright excitons}\label{sec:disc_bx}

Figure~\ref{fig:transitions} shows that bright excitons in ML-WSe$_2$ have exceptionally fast energy relaxation. Figure~\ref{fig:transitions}(d) and (e) show the kernel functions of the energy relaxation profile for the transitions $B \rightarrow B$ in hole-doped ML and $B \rightarrow D$ in electron-doped ML, respectively. Integration of the Kernel function over all possible final-state kinetic energies yields the respective rates in Figs.~\ref{fig:transitions}(a) and (b). The salient yellowish bands in the lower parts of Figs.~\ref{fig:transitions}(d) and (e) imply that a bright exciton is likely to lose most of its initial kinetic energy in one exchange scattering event. As a result, bright excitons in electron-doped MLs are more likely to become dark prior to reaching the light cone.  In hole-doped MLs, on the other hand, exchange scattering improves the probability of bright excitons to reach the light cone before  they turn indirect or dark by emitting $K_3$ or $\Gamma_5$ phonons.  This behavior explains the experimental results showing that the PL intensity of bright excitons is strongest when the ML is  hole-doped.

\subsubsection{Indirect excitons}\label{sec:disc_ix}

We now try to unravel the PL properties of indirect excitons. Their elusive nature is rendered through combination of weak optical transition  and ultrafast relaxation to dark species in the presence of free electrons, as shown from the transition $I \rightarrow D$ in Figs.~\ref{fig:transitions}(b) and (c). The only way to see the signature of indirect excitons in the PL is when the ML is nearly charge neutral. Their lifetime can then become dramatically longer since time-reversal symmetry suppresses their phonon-induced transition to dark excitons following spin-flip intervalley scattering of the electron or hole components \cite{Song_PRL13}. Because photoexcitation inevitably introduces electrons in the CB (especially for excitation energies above the band gap), there is the possibility of having a residual density of free electrons through which indirect excitons become dark. The only requirement for driving the transition $I \rightarrow D$ is  that free electrons populate the opposite valley to that of the electron in the indirect exciton. Depletion of electrons from this valley can be achieved by  pumping the ML with a circularly polarized light that excites bright excitons in the $K$ valley. These excitons can then relax to indirect and dark excitons either through phonon emission or exchange scattering with the residual free electrons in the bottommost  valleys ($B \rightarrow I$ and $B \rightarrow D$). The exchange scattering leaves behind free electrons in the top valley of $K$, which preferably relax to the bottommost  valley in $-K$ through spin-conserving intervalley scattering. This continuous process mostly depletes electrons from the bottommost $K$ valley, thereby weakening the exchange-scattering transition $I \rightarrow D$ for excitons whose electron resides in $-K$. As a result, the population of indirect excitons is enhanced.  This behavior is consistent with the experimental findings of Fig.~\ref{fig:I}, in which indirect excitons are seen best when the ML is photoexcited by a circularly polarized light.

In the absence of free electrons, we should consider exciton-exciton  scattering as additional contributor to the observed behavior.  The exchange interaction between two cold indirect excitons turns them to two hot dark excitons, $I^0_K + I^0_{-K} \rightarrow D^0_K + D^0_{-K}$, where the subscript denotes the valley of the hole in the exciton.  For a given total density of indirect excitons, the  scattering rate is optimal when the densities of  $I^0_K$ and $I^0_{-K}$ are equal, which is the case when the photoexcitation in linearly polarized. At low powers and charge-neutral MLs, the exciton-exciton scattering time can still affect the lifetime of indirect excitons, and hence their density, as long as it is faster than their recombination.

\subsection{Trions}\label{sec:disc_t}

\begin{figure}
\includegraphics[width=8.5cm]{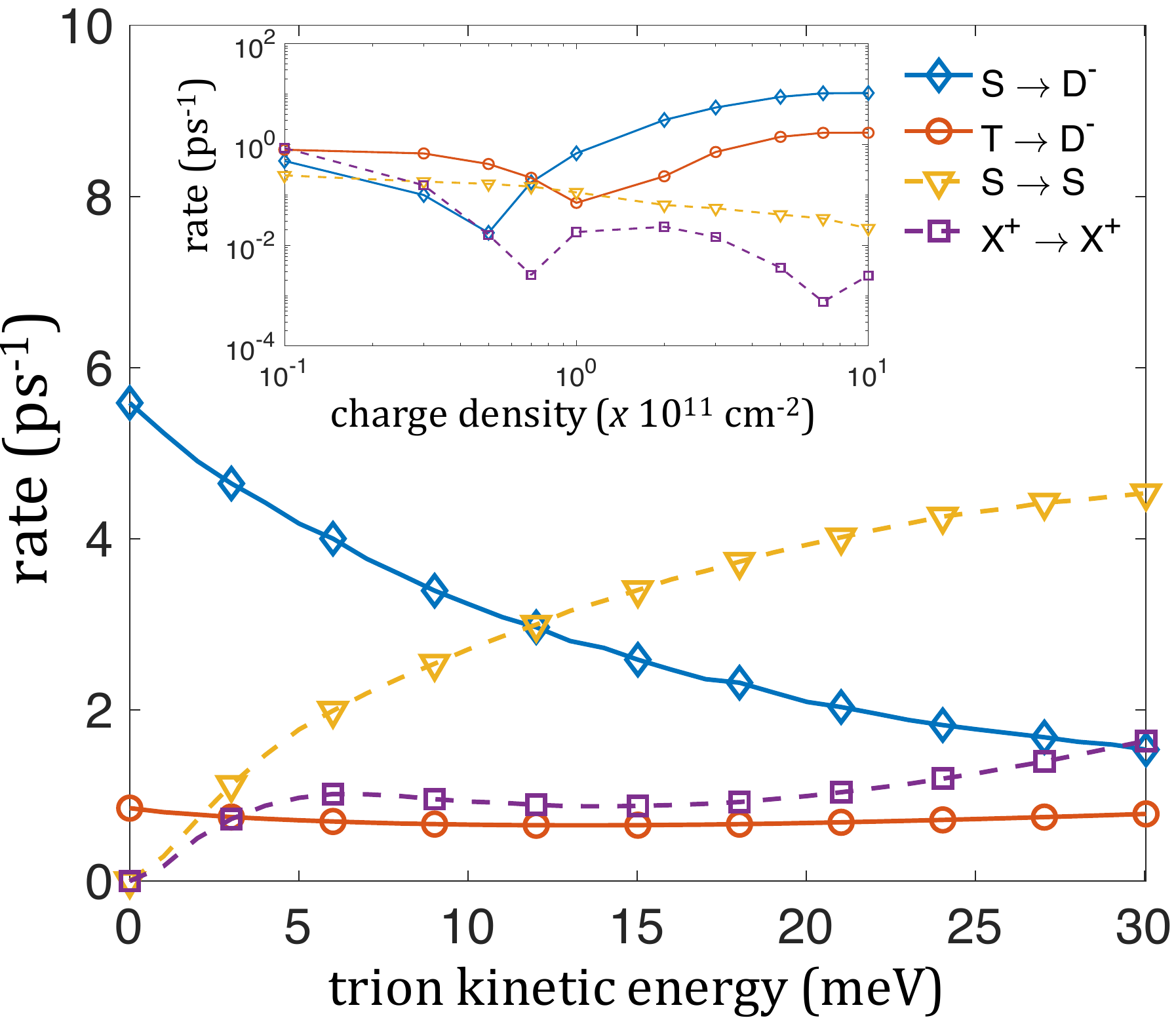}
 \caption{Calculated transition rates between various trion species as a function of the trion initial  kinetic energy in ML-WSe$_2$ at 4$\,$K. The free-charge density is  3$\times$10$^{11}$~cm$^{-2}$. The transitions  $T \rightarrow T$ (not shown) and $S \rightarrow S$ are similar. Inset: The respective rates as a function of charge density for thermal trions ($\varepsilon\,=\,k_BT$).
} \label{fig:ST}
\end{figure}

Similar to the darkening of excitons in electron-doped ML-WSe$_2$, singlet and triplet bright trions become dark if the free electron with which they scatter is from a bottommost valley that is not occupied by the electrons of the bright trion. The final state includes a dark trion with lower energy than the bright trion and free electron in the top valley of the CB (these electrons will eventually thermalize and relax to the bottommost valleys).  The solid lines in Fig.~\ref{fig:ST} indicate that the lifetime of bright negative trions is limited by exchange scattering with  free electrons. The rates are calculated when the free-charge density is 3$\times$10$^{11}$~cm$^{-2}$. The dashed lines in Fig.~\ref{fig:ST} correspond to energy relaxation in which bright trions retain their character, here shown for the positive trion and negative singlet trion (the triplet case is similar).  The formalism used to calculate the transitions $S \rightarrow S$ and $X^+ \rightarrow X^+$ is discussed in Appendix~\ref{app:M}. 

The inset of Fig.~\ref{fig:ST} shows the charge-density dependence of the relaxation rates of thermal trions (kinetic energy is $K_BT$). The non-trivial dependence on charge density is caused by a confluence of two effects. The first one is the restricted phase space for scattering when the kinetic energy of the trion is small. The second effect is due to screening, which becomes less effective when the wavevector of the free charge is changed by more than 2$k_F$ after scattering (see Appendix~\ref{app:details}). The combination of both effects can explain why the transitions $S \rightarrow S$ and   $X^+ \rightarrow X^+$ slightly decrease for thermal trions when the charge density increases: the scattering phase space is restricted to scattering with free-charge near the Fermi surface where the change in wavevector after scattering is small (stronger screening effect). In addition, the two effects explain why the transitions  $S \rightarrow D^-$ and   $T \rightarrow D^-$ eventually increase when the charge density increases:  the energy difference between these trion species requires a relatively large change in wavevector, such that more free particles can take part in the scattering.  Finally, the inset shows that the rates have a dip when the charge density is slightly below 10$^{11}$~cm$^{-2}$. The reason is that the two aforementioned effects result in optimal suppression of the scattering when the kinetic energy of the trions is similar to the Fermi energy. The transition $X^+ \rightarrow X^+$ shows a second dip at higher density, which is attributed to the offset between the direct and exchange scattering terms (see Appendix~\ref{app:M}).

\begin{figure*}
\includegraphics[width=16cm]{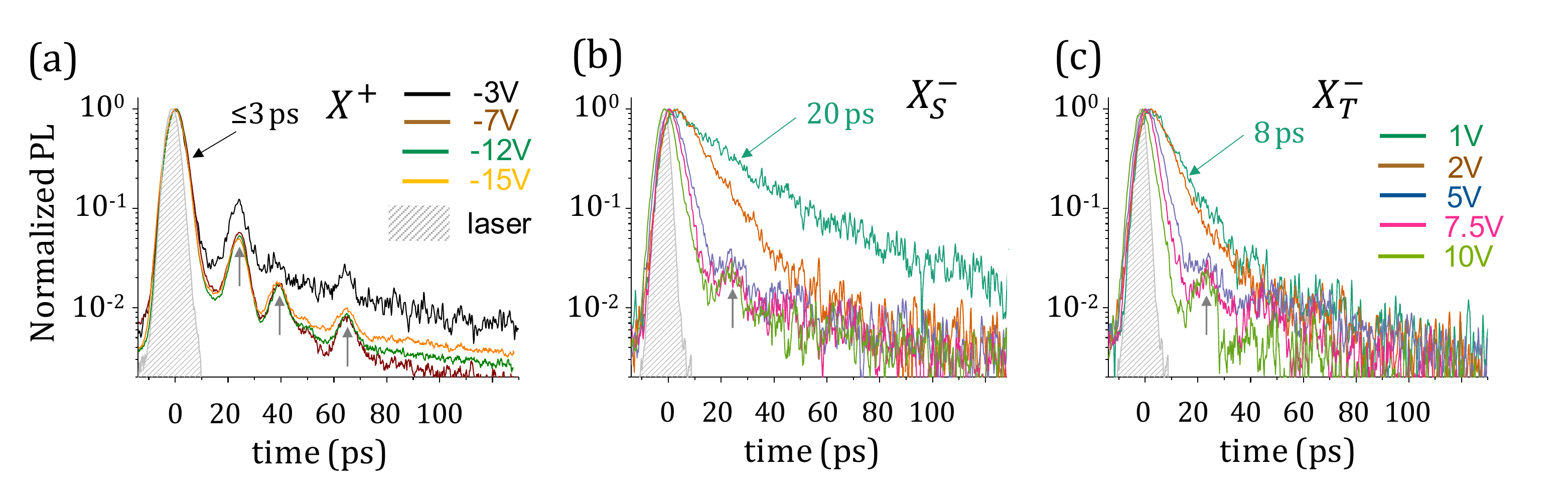}
 \caption{Time-resolved PL measurements of bright trions in ML-WSe$_2$.  The normalized PL of the positive trion is shown in (a), singlet in (b), and triplet in (c). The PL bumps marked by the vertical arrows are artifacts due to laser reflections.} \label{fig:time}
\end{figure*}

To make a direct connection with experiment, we present time-resolved PL measurements from which we extract the trion lifetimes at various charge densities.  Figure~\ref{fig:time} shows time-resolved PL measurements of (a) the positive trion, (b) singlet negative trion, and (c) triplet negative trion. The measured decay time of the positive trion shows no dependence on the gate voltage (hole density), and it is limited by the instrumentation response function (i.e., shorter than 3~ps). The decay time of the singlet (triplet) trion is $\sim$20~ps ($\sim$8~ps) when the gate voltage is 1~V. These timescales are shortened when the gate voltage (electron density) increases, and become instrumentation-limited when the gate voltage is $\gtrsim$5~V. As we explain below, these results correspond well with the discussion we had in Sec.~\ref{sec:ph_rel} and the results of Fig.~\ref{fig:ST}.

Starting with positive bright trions, we have explained in  Sec.~\ref{sec:ph_rel} that their lifetime is governed by the interaction of their electron with zone-edge phonons $K_3$ and zone-center phonons $\Gamma_5$. Thus, we can write that  $\tau_{+}^{-1} \simeq \tau_3^{-1} + \tau_{5}^{-1}$, where $\tau_{3}$ and $\tau_{5}$ are the respective scattering times. Since the phonon-induced scattering times are independent of charge density, we can better understand the time-resolved PL data in Fig.~\ref{fig:time}(a), wherein the lifetime of positive bright trions is independent of hole density. Furthermore, we can use the time-resolved PL results to set an upper limit to the lifetime of positive bright trions, $\tau_{+} \lesssim$ 3~ps. 

The lifetime of bright-negative triplet trions is governed both by exchange scattering, as shown in Fig.~\ref{fig:ST}, and the electron-phonon interaction with zone-center phonons, as we have discussed in Sec.~\ref{sec:ph_rel}. Thus, we can write that $\tau_{T}^{-1} \simeq  \tau_x^{-1} + \tau_5^{-1}$. The exchange rate ($\tau_x^{-1}$) depends on the density of free electrons in the bottommost CB valley that is not occupied by the triplet's electrons. Figure~\ref{fig:time}(c) shows that the lifetime of the triplet trion starts from $\sim$8~ps at 1~V and it becomes shorter than the  instrumentation response time (3~ps) when the gate voltage is $\gtrsim$5~V.  Thus, we assume that $\tau_5 \sim 8$~ps. Recalling the previous result that $\tau_{+}^{-1} \simeq \tau_3^{-1} + \tau_5^{-1} \lesssim 3$~ps$^{-1}$, we can further conclude that $\tau_3$ is about three times or more shorter than $\tau_5$. This conclusion agrees with the findings of He \textit{et al.} who have shown that spin-conserving intervalley relaxation, mediated by zone-edge phonons ($K_3$), is stronger than spin-flip intravalley relaxation that is mediated by zone-center phonons ($\Gamma_5$) \cite{He_NatComm20}.

Turning to bright-negative singlet trions, we have seen from energy considerations in Sec.~\ref{sec:ph_rel} that there should be a phonon bottleneck that limits their relaxation to dark trions (i.e., the phonon energy of $K_3$ is too large to facilitate the transition of a cold singlet to dark one). Thus, we can write that $\tau_{S}^{-1} \simeq  \tau_x^{-1}$, where the rate depends on the density of free electrons in the bottommost CB valley that is not occupied by the singlet's electrons. According to Fig.~\ref{fig:time}(b), the lifetime of the singlet trion starts from $\sim$20~ps at 1~V and it becomes shorter than the  instrumentation response time (3~ps) when the gate voltage is $\gtrsim$5~V.  As we have mentioned in Sec.~\ref{sec:ph_rel}, the lifetime of the singlet trion may also have some contribution from zone-edge phonon modes whose energy is small enough to allow the transition to dark trions (such as $K_2$). Because selection rules indicate a much weaker interaction with these modes compared with $K_3$ (and hence slower), we can better understand why the lifetime of the singlet trion we observed at 1~V is the longest.

\subsubsection{Relative PL intensities of trions}\label{sec:disc_rel}

In experiments with continuous-wave photoexcitation, the PL intensity of a certain excitonic complex is commensurate with its steady-state population, which in turn is commensurate with the lifetime of the complex. Trying to understand the relative PL intensities of bright trions, we need not only consider how quickly they become dark (i.e., what is their lifetime), but also how effective is their formation. Exchange scattering alone cannot explain the interplay between the formation and lifetime of the trions. For example, the PL intensity of bright negative trions is strongest when the laser excites free electron-hole pairs in the continuum, as can be seen by comparing the PL intensities of $X^-_{S/T}$ in Fig.~\ref{fig:various}(a) and (d) versus the ones in Figs.~\ref{fig:various}(b)-(c) and (e)-(f). The reason for this behavior is not clear, and at this point we can only conjecture that it has to do with efficient trimolecular trion formation in which free electron-hole pairs bind to free electrons \cite{Portella-Oberli_PRL09}. In all other cases, the binding is between bound bright excitons and free electrons. The study of trion formation following off-resonance photoexcitation is beyond the scope of this paper, and below we attempt to clarify the shoot-up in the PL intensity of bright negative trions by focusing on the change in their lifetime. 

Figures~\ref{fig:time}(b) and (c) show that the singlet and triplet trions lifetimes are relatively long when the electron density is small. This observation corresponds well with the weaker exchange  scattering that darken the trions at small electron densities, as shown in Fig.~\ref{fig:ST}. In addition to the dependence on electron density, the relative PL intensities of the triplet and singlet trions, $X^-_T$ and $X^-_S$,  depend on the laser polarization. Figure~\ref{fig:various}(a) shows that the dominant peak is $X^-_S$ when the ML is photoexcited by a linearly polarized light.  This dominance is understood from the longest lifetime of the singlet trion when the electron density is small (Fig.~\ref{fig:time}). Figure~\ref{fig:map} shows that the dominant peak is $X^-_T$ when the photoexcitation is circularly polarized. This behavior is driven by the dynamical valley polarization effect, wherein photoexcitation of bright excitons in the $K$ valley leads to accumulation (depletion) of free electrons  in the bottommost CB of the $-K$ ($K$) valley \cite{Robert_NatComm21,Hao_arXiv21}. As a result, the triplet formation is more effective because there are more free electrons from $-K$ that bright excitons in $K$ can bind with.  Equally important, the PL intensity of triplet trions is enhanced when their lifetime is enhanced. The latter is caused by the depletion of electrons from the $K$ valley, which in turn suppresses the triplet-to-dark transition due to exchange scattering. A similar lifetime enhancement effect was discussed in the context of indirect excitons in Sec.~\ref{sec:disc_ix}.  

Turning to the PL intensity of positive trions, Figs.~\ref{fig:map} and \ref{fig:various} show a gradual increase in the intensity of $X^+$ when the hole density increases. This behavior can be explained through the interplay between the formation efficiency and lifetime of positive trions. The  formation process requires that free holes bind to bright excitons before they turn dark. In the presence of holes, the darkening of bright excitons  through exchange scattering is ineffective, and they can only  become dark through phonon emission ($K_3$ or $\Gamma_5$).  As was shown in Fig.~\ref{fig:time}(a), the lifetime of positive trions does not depend on the density of free holes. On the other hand,  the trion formation time is by definition faster when the density of free holes increases. The outcome is a gradual increase in the PL intensity of $X^+$ when the density of free holes increases. The PL intensity eventually ceases to grow when the formation time of positive trions is much faster than the darkening time of bright electron-hole pairs (i.e., when all excited electron-hole pairs become positive trions).  

\subsection{Epilogue}

Before we conclude this work, three points regarding the previous analyses merit discussion. The first one deals with the relation between charge density and gate voltage. We have mentioned that a change of 1~V in our devices amounts to a change of $10^{11}$~cm$^{-2}$ in the density of electrons (or holes). This correspondence is based on a simple capacitance model \cite{Robert_PRL21,Robert_NatComm21}. Yet, it is only true when the ML is not photoexcited. Otherwise, the charge density includes a contribution from charged complexes (mostly dark trions). In other words, the density of free electrons or holes in the photoexcited ML is smaller than the one we would have sans photoexcitation. Thus, care should be taken when one wishes to compare the calculated results at a given density of free electrons with the experimental results at a given voltage level. 

The second point deals with our assumption that the phonon-induced scattering times are similar for all excitonic complexes. Namely, we have assumed similar scattering times when excitons, positive trions or negative trions interact with zone-edge phonons $K_3$ and zone-center phonons $\Gamma_5$. The justification for this approximation is twofold. Firstly,  selection rules are such that only electrons in the CB can interact with these phonon modes. Secondly, the short-range nature of the electron-phonon interaction with $K_3$ and $\Gamma_5$ means that holes are spectators and the scattering amplitudes are similar regardless of whether we deal with free electrons, excitons, or trions. 

The third and last point is that our calculations reveal exchange scattering rates that are too fast  compared with the observed behavior in experiment. For example, the measured singlet lifetime is 20~ps  at 1~V, whereas the calculated exchange-scattering lifetime is an order of magnitude shorter when the charge density is $\sim$10$^{11}$~cm$^{-2}$. As mentioned in the first point, the electron density may be evidently smaller than $\sim$10$^{11}$~cm$^{-2}$ at 1~V when the laser pulse introduces enough electron-hole pairs to convert many of the free electrons to trions.  We also emphasize that the Coulomb potential we are using in this work is the statically screened one (Appendix~\ref{app:details}). Therefore, the discrepancy cannot be attributed to an overestimated scattering potential. 

With all these caveats taken into account, exchange scattering still provides a compelling overarching argument to explain many of the observed phenomena we see in experiment. And when augmenting the exchange scattering mechanism with exciton-phonon interactions ($K_3$ and $\Gamma_5$), we get a self-consistent description of the physics. 

\section{Conclusion}\label{sec:conc}

We have studied the key role played by exciton-electron and trion-electron exchange scattering. Whereas the signature of phonons can be readily recognized from phonon replicas of dark and indirect excitonic complexes in the emission spectrum, the signature of exchange scattering is subtle.  The latter effect can be recognized from the intensity dependence of various optical transitions on the charge density in the monolayer. 

We have shown that exchange scattering in monolayer WSe$_2$ manifests through darkening of bright excitons when the monolayer is lightly doped with electrons and through darkening of bright trions when the monolayer is moderately doped with electrons.  On the other hand, the exchange scattering does not darken the bright exciton when the monolayer is doped with holes. Instead of darkening, the scattering with free holes enhances the energy relaxation of bright excitons, pushing them closer to the light cone before they can become dark or indirect by phonon emission. In addition, the exchange scattering mechanism helps us to better understand the fast rise of the photoluminescence intensity of bright trions as soon as electrons are added to the monolayer. The trions' darkening is suppressed at small electron densities, and therefore, their lifetime is enhanced. The effect is lost when the density of free electrons continues to increase and the bright trions turn dark through exchange scattering.

The findings and conclusions of this work are not model dependent. Namely, one would reach at similar conclusions either by using a simple 2D hydrogen wavefunctions and potential or accurate numerical techniques. The important two factors to keep in mind are the relatively small dielectric screening, which leads to stronger scattering potential compared with conventional quantum well systems. And most importantly, the reversed order of the conduction-band valleys in tungsten-based monolayers. The result of which is that dark excitonic complexes comprise thermal electrons (i.e., from the bottommost valleys in the conduction band). Together with the fact that dark and indirect excitonic complexes lie lower in energy than bright ones, we end up with the exchange-driven darkening effect. In this view, the results we got in this work for ML-WSe$_2$ should be similar in ML-WS$_2$, wherein the spin-valley configuration of the conduction band is the same.

To gain further understanding of optical properties in monolayer semiconductors,  future experiments and theoretical studies should provide better understanding of the formation of bright excitonic complexes as a function of laser power, excitation wavelength, and electrostatic doping. Such understanding will shed light on the interplay between the formation efficiency of bright complexes and their lifetime, from which we can quantitatively model the photoluminescence  intensity. To complete the picture, we additionally need to better understand the intrinsic types of non-radiative recombination processes that govern the lifetime of dark excitons and trions.  

\acknowledgments{The work in Rochester on excitons was supported by the Department of Energy, Basic Energy Sciences, under Contract No. DE-SC0014349, and the work on trions by the Office of Naval Research, under Contract No. N000142112448. Cedric Robert and Xavier Marie acknowledge funding from ANR 2D-vdW-Spin, ANR ATOEMS and ANR MagicValley. Xavier Marie also acknowledges the Institut Universitaire de France.}

\appendix
\section{Technical Details of the calculations}\label{app:details}

The exciton-electron and trion-electron matrix elements are calculated using the statically screened potential due to the presence of free electrons. We have used the following statically screened potential to calculate the exciton wavefunctions and matrix elements \cite{Scharf_JPCM19}, 
\begin{eqnarray} \label{eq:Vq}
 V_{\mathbf{q}} = \frac{2\pi e^2}{A\epsilon_d(q)} \cdot \frac{1}{ q + \kappa(q)} \,\,.
\end{eqnarray}
$A$ is the monolayer area and $\epsilon_d(q)$ is the dielectric screening function, where its $q$-dependence stems from non-local effects \cite{Rytova_PMPA67,Keldysh_JETP79,Schmitt-Rink_JL85}.  In this work we have followed the model in Ref.~\cite{VanTuan_PRB18} for the dielectric function, but this choice is not critical and the results should be qualitatively similar when choosing other models \cite{Cudazzo_PRB11,Latini_PRB15,Qiu_PRB16}. Using the random-phase approximation (RPA), the screening wavenumber is \cite{Ando_RMP82} 
\begin{equation}\label{eq:kappa_RPA}
\kappa(q)=\frac{g_sg_v e^2 m_\mathrm{b}}{\hbar^2\epsilon_d(q)} \left[1-\sqrt{1-\left( \frac{2k_\mathrm{F}}{q}\right)^2}\Theta(q-2k_\mathrm{F})\right]\,.\,\,\,\,\,\,
\end{equation}
$g_s=1$ and $g_v=2$ are the spin and valley degeneracies, respectively. $m_b$ is the effective mass of an electron in the bottommost CB valley. In hole-doped systems, this mass is replaced with that of the topmost VB valley.  $k_F$ is the Fermi wavenumber related to the charge density through $2\pi n =k_F^2$. Omitting the term in square brackets on the right-hand side of Eq.~(\ref{eq:kappa_RPA}) amounts to the Thomas-Fermi approximation of an ideal 2D system. However, this approximation overestimates the screening, leading to unphysical results where the screening length is independent of the charge density. The static RPA model resolves this problem,  where the screening is suppressed when the wavenumber is larger than $2k_F$. 

\section{Wavefunctions}\label{app:trion_direct}
We have used the stochastic variational method (SVM) to express excitonic wavefunctions in terms of correlated Gaussians \cite{Varga_CPC08,Varga_PRC95,Mitroy_RMP13,Kidd_PRB16,Donck_PRB17,VanTuan_PRL19}.  We have followed the procedure in Refs.~\cite{Yang_PRB20,VanTuan_PRB18}, where the wavefunctions of neutral excitons have the form
\begin{eqnarray} 
\varphi(r) &=& \sum_{j=1}^n C_j \exp\left( -\tfrac{1}{2}\alpha_{j} r^2     \right) \,\,.\label{eq:svm_form}
\end{eqnarray} 
The number of correlated Gaussians needed to accurately describe the ground state is $n$, and $r$ is the distance between the electron and hole.  Using this wavefunction form, we can perform the integration over $\mathbf{r}$ analytically. That is, the matrix elements become a discrete sum over elements that are expressed in terms of the (real) variational parameters, $C_j$ and $\alpha_{j}$. Given that it is sufficient to use a few tens of correlated Gaussians to accurately describe the exciton states, the calculation is efficient and fast. Wavefunctions of trions have the form
\begin{eqnarray} 
\varphi(\mathbf{r}_1,\mathbf{r}_2) &=& \sum_{i=1}^N C_j \exp\left( -\alpha_{i} r_1^2  - \beta_{i} r_2^2   - \gamma_{i} \mathbf{r}_1 \cdot \mathbf{r}_2 \right) ,\,\,\,\,\,\,\,\,\,\,\label{eq:svm_trion_form}
\end{eqnarray} 
where $\mathbf{r}_1$ and $\mathbf{r}_2$ are the distances of the first and second electrons from the hole for negative trions (and vice versa for positive trions).  $C_i$ $\alpha_{i}$, $\beta_{i}$ and $\gamma_{i}$ are real variational parameters.

\section{Matrix elements}\label{app:M}
The derivation of the matrix elements in Eqs.~(\ref{eq:Vee_eh}) and (\ref{eq:SD}) follows the work of Ramon, Mann and Cohen \cite{Ramon_PRB03}. The changes in this work are the consideration of various excitonic species before and after scattering, inclusion of static screening in the scattering Coulomb potential, and the use of exciton and trion wavefunctions from SVM instead of using hydrogen-type wavefunctions.

Figure~\ref{fig:ST} includes calculations of the intra-species scattering rates between singlet trions ($S \rightarrow S$) as well as between positive trions $X^+ \rightarrow X^+$. The former is calculated assuming that the singlet trion scatters with an electron from the bottommost CB valley that is not occupied by the singlet electrons. This way, we can compare the direct and exchange terms on equal footings (i.e., $S \rightarrow S$ vs $S \rightarrow D^-$). The matrix element for the transition $S \rightarrow D^-$ is provided in Eq.~(\ref{eq:SD}). The direct matrix element for the transition $S \rightarrow S$ follows,
\begin{widetext}
\begin{eqnarray} \label{eq:StoS}
\!\!\!\!\!\!&\!\!\!\!\!\!&\!\!\!\!\!\! \!\!\! \left\langle  X^-_S, \mathbf{k}_T \!+\! \mathbf{q} \,\,;\,\, b, \mathbf{k}_e\!-\!\mathbf{q} \,\left|\,  V_{eh} + V_{ee} \, \right| \,  X^-_S, \mathbf{k}_T\,;\,b , \mathbf{k}_e  \right\rangle   = \frac{V_{\mathbf{q}}}{A}\sum_{\mathbf{k},\mathbf{p}}   \psi_S^\ast(  \mathbf{k} \,,\,  \mathbf{p})         \Big\{  \nonumber  \\  && \qquad \qquad \qquad  \,\,\,\, \psi_S(   \mathbf{k}  - \bar{\alpha}_{t} \mathbf{q}\,,\,  \mathbf{p} +\alpha_b \mathbf{q})  + \psi_S(   \mathbf{k}  + \alpha_t \mathbf{q}\,,\,  \mathbf{p}  -\bar{\alpha_{b}}  \mathbf{q}) - \psi_S(    \mathbf{k}  + \alpha_t \mathbf{q}\,,\,  \mathbf{p} +\alpha_b \mathbf{q}) \Big\}\,. 
\end{eqnarray}
\end{widetext}
The symbols are defined after Eq.~(\ref{eq:SD}) and we added that $\bar{\alpha_i} = 1 - \alpha_i$. The matrix element for the transition $X^+ \rightarrow X^+$ is different because both direct and exchange terms leaves the trion at the same state (i.e., $X^+$ remains $X^+$ after scattering with a free hole). Repeating the analysis for this case, the exchange matrix element is similar to Eq.~(\ref{eq:SD}) and the direct one is similar to Eq.~(\ref{eq:StoS}). The needed changes are $\psi_{S/D} \rightarrow \psi_{X^+}$, $\alpha_i \rightarrow \alpha_h = m_h/M_{X^+}$ for $i=\{b,t,D\}$, $V_{ee} \rightarrow V_{hh}=V_{hh_1} + V_{hh_2}$, and the free holes are from the topmost VB valley ($j=t$, and $\mathbf{k}_e \rightarrow \mathbf{k}_h$).  The exchange and direct matrix elements of dark trion transitions, $D^{\pm} \rightarrow D^{\pm}$, are similar and require the appropriate  parameter changes. 

\section{Compiled list of parameters}\label{app:params}

The following parameters are used for ML-WSe$_2$.

\begin{enumerate}

\vspace{-1mm}
\item The effective masses are $m_{t}=0.29m_0$ (top valley of the CB), $m_{b}=0.4m_0$ (bottom valley  of the CB), and $m_h=0.36m_0$ (top valley of the VB) \cite{Kormanyos_2DMater15}. Using these values, the translational exciton masses are $M_{X^0}=0.65m_0$ and $M_{D^0}=M_{I^0}=0.76m_0$. Similarly,  the translational trion masses are $M_{X^+}=1.01m_0$, $M_{X^-_S}=M_{X^-_T}=1.05m_0$, $M_{D^+}=1.12m_0$ and $M_{D^-}=1.16m_0$. The kinetic energies of electrons, holes, excitons and trions are evaluated by parabolic energy dispersion.

\vspace{-1mm}
\item  The non-local dielectric function, $\epsilon_d(q)$, and its parameters for hBN encapsulated ML-WSe$_2$ are the same as in Ref.~\cite{VanTuan_PRB18}. 

\vspace{-1mm}
\item The energy values used in the calculations follow empirical results. The spin splitting energy of the CB is $\Delta_C\,=\,14$~meV \cite{Kapuscinski_NatComm21}.  The difference in energy resonances of the exciton species are $\Delta_{\text{bd}}\,=\,40$~meV and $\Delta_{\text{bi}}\,=\,31$~meV. The resonance energy of the negative singlet (triplet) trion is 19~(26)~meV above that of the dark trion.  The resonance energy of the positive trion is 33~meV above that of the dark one.

\end{enumerate}


\begin{thebibliography}{99}


\bibitem{Wang_RMP18}  G. Wang, A. Chernikov, M. M. Glazov, T. F. Heinz, X. Marie, T. Amand, and B. Urbaszek, Colloquium: Excitons in atomically thin transition metal dichalcogenides, Rev. Mod. Phys. \textbf{90}, 021001 (2018).
\bibitem{Mak_NatPhoton18} K. F. Mak, D. Xiao, and J. Shan, Light–valley interactions in 2D semiconductors, Nat. Photon. \textbf{12}, 451 (2018).

\bibitem{Jones_NatPhys16} A. M. Jones, H. Yu, J. Schaibley, J. Yan, D. G. Mandrus, T. Taniguchi, K. Watanabe, H. Dery, W. Yao, and X. Xu, Excitonic luminescence upconversion in a two-dimensional semiconductor, Nat. Phys. \textbf{12}, 323 (2016).
\bibitem{Courtade_PRB17} E. Courtade, M. Semina, M. Manca, M. M. Glazov, C. Robert, F. Cadiz, G. Wang, T. Taniguchi, K. Watanabe, M. Pierre, W. Escoffier, E. L. Ivchenko, P. Renucci, X. Marie, T. Amand, and B. Urbaszek, Charged excitons in monolayer WSe$_2$: experiment and theory, Phys. Rev. B \textbf{96}, 085302 (2017).

\bibitem{Zhang_NatNano17} X.-X. Zhang, T. Cao, Z. Lu, Y.-C. Lin, F. Zhang, Y. Wang, Z. Li, J. C. Hone, J. A. Robinson, D. Smirnov, S. G. Louie, and T. F. Heinz, Magnetic brightening and control of dark excitons in monolayer WSe$_2$, Nat. Nanotechnol. \textbf{12}, 883 (2017).
\bibitem{Zhou_NatNano17} Y. Zhou, G. Scuri, D. S. Wild, A. A. High, A. Dibos, L. A. Jauregui, C. Shu, K. De Greve, K. Pistunova, A. Y. Joe, T. Taniguchi, K. Watanabe, P. Kim, M. D. Lukin, and H. Park, Probing dark excitons in atomically thin semiconductors via near-field coupling to surface plasmon polaritons, Nat. Nanotechnol. \textbf{12}, 856 (2017).
\bibitem{Wang_PRL17} G. Wang, C. Robert, M. M. Glazov, F. Cadiz, E. Courtade, T. Amand, D. Lagarde, T. Taniguchi, K. Watanabe, B. Urbaszek, and X. Marie,  In-plane propagation of light in transition metal dichalcogenide monolayers: optical selection rules, Phys. Rev. Lett. \textbf{119}, 047401 (2017).
\bibitem{Chen_NatComm18} S.-Y. Chen, T. Goldstein, T. Taniguchi, K. Watanabe, and J. Yan,  Coulomb-bound four- and five-particle intervalley states in an atomically-thin semiconductor, Nat. Commun. {\bf 9}, 3717 (2018).
\bibitem{Ye_NatComm18} Z. Ye, L. Waldecker, E. Y. Ma, D. Rhodes, A. Antony, B. Kim, X.-X. Zhang, M. Deng, Y. Jiang, Z. Lu, D. Smirnov, K. Watanabe, T. Taniguchi, J. Hone, and  T. F. Heinz, Efficient generation of neutral and charged biexcitons in encapsulated WSe$_2$ monolayers,  Nat. Commun. {\bf 9}, 3718 (2018).
\bibitem{Li_NatComm18} Z. Li, T. Wang, Z. Lu, C. Jin, Y. Chen, Y. Meng, Z. Lian, T. Taniguchi, K. Watanabe, S. Zhang, D. Smirnov, and S.-F. Shi,  Revealing the biexciton and trion-exciton complexes in BN encapsulated WSe$_2$, Nat. Commun. {\bf 9}, 3719 (2018).  
\bibitem{Barbone_NatComm18} M. Barbone, A. R.-P. Montblanch, D. M. Kara, C. Palacios-Berraquero, A. R. Cadore, D. De Fazio, B. Pingault, E. Mostaani, H. Li, B. Chen, K. Watanabe, T. Taniguchi, S. Tongay, G. Wang, A. C. Ferrari, and M. Atat\"ure ,  Charge-tuneable biexciton complexes in monolayer WSe$_2$, Nat. Commun. {\bf 9}, 3721 (2018).

\bibitem{He_NatComm20} M. He, P. Rivera, D. V. Tuan, N. P. Wilson, M. Yang, T. Taniguchi, K. Watanabe, J. Yan, D. G. Mandrus, H. Yu, H. Dery, W. Yao, and X. Xu, Valley phonons and exciton complexes in a monolayer semiconductor, Nat. Commun. \textbf{11}, 618 (2020).
\bibitem{Liu_PRL20} E. Liu, J. van Baren, C.-T. Liang, T. Taniguchi, K. Watanabe, N. M. Gabor, Y.-C. Chang, and C.-H. Lui, Multipath optical recombination of intervalley dark excitons and trions in monolayer WSe$_2$, Phys. Rev. Lett. \textbf{124}, 196802 (2020).
\bibitem{Li_ACS19} Z. Li, T. Wang, C. Jin, Z. Lu, Z. Lian, Y. Meng, M. Blei, M. Gao, T. Taniguchi, K. Watanabe, T. Ren, T. Cao, S. Tongay, D. Smirnov, L. Zhang, S.-F. Shi, Momentum-dark intervalley exciton in monolayer tungsten diselenide brightened via chiral phonon, ACS Nano \textbf{13}, 14107 (2019). 
\bibitem{Tang_NatComm19} Y. Tang, K. F. Mak, and J. Shan, Long valley lifetime of dark excitons in single-layer WSe$_2$, Nat. Commun. \textbf{10}, 4047 (2019). 

\bibitem{Dery_PRB15} H. Dery and Y. Song, Polarization analysis of excitons in monolayer and bilayer transition-metal dichalcogenides,  Phys. Rev. B \textbf{92}, 125431 (2015). 
\bibitem{Slobodeniuk_2DMater16} A. O. Slobodeniuk and D. M. Basko, Spin-flip processes and radiative decay of dark intravalley excitons in transition metal dichalcogenide monolayers, 2D Mater. \textbf{3}, 035009 (2016).
\bibitem{Scharf_JPCM19} B. Scharf, D. Van Tuan, I. \v{Z}uti\'c, and H. Dery, Dynamical screening in monolayer transition-metal dichalcogenides and its manifestations in the exciton spectrum, J. of Phys.: Conden. Mater. \textbf{31}, 203001 (2019).


\bibitem{Honold_PRB89} A. Honold, L. Schultheis, J. Kuhl, and C. W. Tu, Collision broadening of two-dimensional excitons in a GaAs single quantum well, Phys. Rev. B \textbf{40}, 6442(R) (1989).
\bibitem{Eccleston_PRB91} R. Eccleston, R. Strobel, W. W. Rühle, J. Kuhl, B. F. Feuerbacher, and K. Ploog, Exciton dynamics in a GaAs quantum well, Phys. Rev. B \textbf{44}, 1395(R) (1991).
\bibitem{Kochereshko_PRB98} V. P. Kochereshko, D. R. Yakovlev, R. A. Suris, W. Ossau, A. Waag, G. Landwehr, P. C. M. Christianen, and J. C. Maan, Combined exciton-electron excitation in quantum wells with a two-dimensional electron gas of low density, Superlattices and Microstructures, \textbf{23}, 283 (1998). 
\bibitem{Ramon_PRB03} G. Ramon, A. Mann, and E. Cohen, Theory of neutral and charged exciton scattering with electrons in semiconductor quantum wells, Phys. Rev. B \textbf{67}, 045323 (2003).
\bibitem{Shahnazaryan_PRB17} V. Shahnazaryan, I. Iorsh, I. A. Shelykh, and O. Kyriienko, Exciton-exciton interaction in transition-metal dichalcogenide monolayers, Phys. Rev. B \textbf{96}, 115409 (2017).
\bibitem{Liu_PRL19} E. Liu, J. van Baren, Z. Lu, M. M. Altaiary, T. Taniguchi, K. Watanabe, D. Smirnov, and C. H. Lui, Gate tunable dark trions in monolayer WSe$_2$, Phys. Rev. Lett. \textbf{123}, 027401 (2019).
\bibitem{Li_NanoLett19} Z. Li, T. Wang, Z. Lu, M. Khatoniar, Z. Lian, Y. Meng, M. Blei, T. Taniguchi, K. Watanabe, S. A. McGill, S. Tongay, V. M. Menon, D. Smirnov, and S.-F. Shi, Direct observation of gate-tunable dark trions in monolayer WSe$_2$, Nano Lett. \textbf{19}, 6886 (2019). 
\bibitem{Robert_NatComm21} C. Robert, S. Park, F. Cadiz, L. Lombez, L. Ren, H. Tornatzky, A. Rowe, D. Paget, F. Sirotti, M. Yang, D. V. Tuan, T. Taniguchi, B. Urbaszek, K. Watanabe, T. Amand, H. Dery, and X. Marie, Spin/Valley pumping of resident electrons in WSe$_2$ and WS$_2$ monolayers, Nat. Commun. \textbf{12}, 5455 (2021).

\bibitem{Stier_PRL18} A. V. Stier, N. P. Wilson, K. A. Velizhanin, J. Kono, X. Xu, and S. A. Crooker, Magnetooptics of exciton Rydberg states in a monolayer semiconductor, Phys. Rev. Lett. \textbf{120}, 057405 (2018).
\bibitem{Liu_PRB19} E. Liu, J. van Baren, T.i Taniguchi, K. Watanabe, Y.-C. Chang, and C.-H. Lui Magnetophotoluminescence of exciton Rydberg states in monolayer 
WSe$_2$, Phys. Rev. B \textbf{99}, 205420 (2019).
\bibitem{Goryca_NatComm19} M. Goryca, J. Li, A. V. Stier, T. Taniguchi, K. Watanabe, E. Courtade, S. Shree, C. Robert, B. Urbaszek, X. Marie, and S. A. Crooker, Revealing exciton masses and dielectric properties of monolayer semiconductors with high magnetic fields, Nat. Commun. \textbf{10}, 4172 (2019).
\bibitem{Kapuscinski_NatComm21} P. Kapu\'{s}ci\'{n}ski, A. Delhomme, D. Vaclavkova, A. O. Slobodeniuk, M. Grzeszczyk, M. Bartos, K. Watanabe, T. Taniguchi, C. Faugeras, and M. Potemski, Rydberg series of dark excitons and the conduction band spin-orbit splitting in monolayer WSe$_2$, Nat. Commun. \textbf{4}, 186 (2021).


\bibitem{Hichri_PRB20} A. Hichri and S. Jaziri, Trion fine structure and anomalous Hall effect in monolayer transition metal dichalcogenides, Phys. Rev. B \textbf{102}, 085407 (2020).
\bibitem{Glazov_JCP20}  M. M. Glazov, Optical properties of charged excitons in two-dimensional semiconductors, J. Chem. Phys. \textbf{153}, 034703 (2020).

\bibitem{Kormanyos_2DMater15}   A. Korm\'{a}nyos, G. Burkard, M. Gmitra, J. Fabian, V. Z\'{o}lyomi, N. D. Drummond, and V. Fal'ko, $\mathbf{k \cdot p}$ theory for two-dimensional transition metal dichalcogenide semiconductors, 2D Mater. \textbf{2}, 022001 (2015). 

\bibitem{VanTuan_PRB18} D. Van Tuan, M. Yang,  and H. Dery, Coulomb interaction in monolayer transition-metal dichalcogenides,  Phys. Rev. B \textbf{98}, 125308 (2018).

\bibitem{Cardona_book} P. Y. Yu and M. Cardona, Fundamentals of Semiconductors, 3rd ed. (Springer, Berlin, 2005).

\bibitem{Qiu_PRL15} D. Y. Qiu, T. Cao, and S. G. Louie, Nonanalyticity, valley quantum phases, and light-like exciton dispersion in monolayer transition metal dichalcogenides: theory and first-principles calculations, Phys. Rev. Lett. \textbf{115}, 176801 (2015).
\bibitem{Echeverry_PRB16} J. P. Echeverry, B. Urbaszek, T. Amand, X. Marie, and I. C. Gerber, Splitting between bright and dark excitons in transition metal dichalcogenide monolayers, Phys. Rev. B \textbf{93}, 121107(R) (2016).
\bibitem{Deilmann_PRB17} T. Deilmann and K. S. Thygesen, Dark excitations in monolayer transition metal dichalcogenides, Phys. Rev. B \textbf{96} 201113 (2017).

\bibitem{Kosmider_PRB13} K. Ko\'{s}mider, J. W. Gonz\'{a}lez, and J. Fern\'{a}ndez-Rossier, Large spin splitting in the conduction band of transition metal dichalcogenide monolayers, Phys. Rev. B \textbf{88}, 245436 (2013).



\bibitem{Wang_NanoLett17} Z. Wang, L. Zhao, K. F. Mak, and J. Shan, Probing the spin-polarized electronic band structure in monolayer transition metal dichalcogenides by optical spectroscopy, Nano Lett. \textbf{17}, 740 (2017).

\bibitem{Danovich_SR17} M. Danovich, V. Z\`{o}lyomi, and V. I. Fal'ko, Dark trions and biexcitons in WS$_2$ and WSe$_2$ made bright by e-e scattering, Sci. Rep. \textbf{7}, 45998 (2017).
\bibitem{Tu_JPCM19} J.-S. Tu, S. Borghardt, D. Gr\"{u}tzmacher, and B. E. Kardyna\l, J. Phys.: Condens. Matter \textbf{31} 415701 (2019).

\bibitem{QE} P. Giannozzi \textit{et al.}, QUANTUM ESPRESSO: a modular and open-source software project for quantum simulations of materials. J. Phys. Condens. Matter \textbf{21}, 395502 (2009).

\bibitem{Song_PRL13} Y. Song and H. Dery, Transport theory of monolayer transition-metal dichalcogenides through symmetry, Phys. Rev. Lett. \textbf{111}, 026601 (2013).

\bibitem{Song_PRB12} Y. Song and H. Dery, Analysis of phonon-induced spin relaxation processes in silicon, Phys. Rev. B \textbf{86}, 085201 (2012).




\bibitem{footnote1} Note that  $\Delta_c$ is absent in these energy considerations because the band gap change  after the transition ($-\Delta_c$) is offset by the respective change in binding energy, $\Delta_{\text{bd}}-\Delta_c$ or $\Delta_{\text{bi}}-\Delta_c$, when excitons relax from bright to dark or indirect species.   






\bibitem{Yang_PRB20} M. Yang, C. Robert, Z. Lu, D. V. Tuan, D. Smirnov, X. Marie, and H. Dery,  Exciton valley depolarization in monolayer transition-metal dichalcogenides,  Phys. Rev. B \textbf{101}, 115307 (2020).

\bibitem{Portella-Oberli_PRL09} M. T. Portella-Oberli, J. Berney, L. Kappei, F. Morier-Genoud, J. Szczytko, and B. Deveaud-Pl\'{e}dran, Dynamics of trion formation in In$_x$Ga$_{1-x}$As quantum wells, Phys. Rev. Lett. \textbf{102}, 096402 (2009).

\bibitem{Hao_arXiv21} K. Hao, R. Shreiner, A. A. High, Optically controllable magnetism in atomically thin semiconductors, arXiv:2108.05931 .


\bibitem{Robert_PRL21} C. Robert, H. Dery, L. Ren, D. Van Tuan, E. Courtade, M. Yang, B. Urbaszek, D. Lagarde, K. Watanabe, T. Taniguchi, T. Amand, and X. Marie, Measurement of conduction and
valence bands $g$-factors in a transition metal dichalcogenide monolayer, Phys. Rev.
Lett. \textbf{126}, 067403 (2021).

\bibitem{Rytova_PMPA67} N. S. Rytova, Screened potential of a point charge in a thin film, Proc. MSU, Phys. Astron. \textbf{3}, 30 (1967).
\bibitem{Keldysh_JETP79} L. V. Keldysh, Coulomb interaction in thin semiconductor and semimetal films, JETP Lett. \textbf{29}, 658 (1979).
\bibitem{Schmitt-Rink_JL85} S. Schmitt-Rink and C. Ell, Excitons and electron-hole plasma in quasi-two-dimensional systems, J. Lumin \textbf{30}, 585 (1985).
\bibitem{Cudazzo_PRB11} P. Cudazzo, I. V. Tokatly, and A. Rubio, Dielectric screening in two-dimensional insulators: Implications for excitonic and impurity states in graphane, Phys. Rev. B \textbf{84}, 085406 (2011). %
%
\bibitem{Latini_PRB15} S. Latini, T. Olsen, and K. S. Thygesen, Excitons in van der Waals heterostructures: The important role of dielectric screening, Phys. Rev. B \textbf{92}, 245123 (2015).
\bibitem{Qiu_PRB16}  D. Y. Qiu, F. H. da Jornada, and S. G. Louie, Screening and many-body effects in two-dimensional crystals: Monolayer MoS$_2$, Phys. Rev. B \textbf{93}, 235435 (2016).

\bibitem{Ando_RMP82} T. Ando, A. B. Fowler, and F. Stern, Electronic properties of two-dimensional systems, Rev. Mod. Phys. \textbf{54} 437 (1982).

\bibitem{Varga_PRC95} K. Varga and  Y. Suzuki, Precise solution of few-body problems with the stochastic variational method on a correlated Gaussian basis, Phys. Rev. C {\bf 52}, 2885 (1995).
\bibitem{Varga_CPC08} K. Varga, Solution of few-body problems with the stochastic variational method II: Two-dimensional systems, Comp. Phys. Comm. {\bf 179}, 591 (2008). 
\bibitem{Mitroy_RMP13} J. Mitroy, S. Bubin, W. Horiuchi, Y. Suzuki, L. Adamowicz, W. Cencek, K. Szalewicz, J. Komasa, D. Blume, and K. Varga, Theory and application of explicitly correlated Gaussians, Rev. Mod. Phys. \textbf{85}, 693 (2013).
\bibitem{Kidd_PRB16} D. W. Kidd, D. K. Zhang, and K. Varga, Binding energies and structures of two-dimensional excitonic complexes in transition metal dichalcogenides, Phys. Rev. B \textbf{93}, 125423 (2016).
\bibitem{Donck_PRB17} M. Van der Donck, M. Zarenia, and F. M. Peeters, Excitons and trions in monolayer transition metal dichalcogenides: A comparative study between the multiband model and the quadratic single-band model, Phys. Rev. B \textbf{96}, 035131 (2017).
\bibitem{VanTuan_PRL19} D. Van Tuan, A. M. Jones, M. Yang, X. Xu, and H. Dery, Virtual trions in the photoluminescence of monolayer transition-metal dichalcogenides,  Phys. Rev. Lett. \textbf{122}, 217401 (2019).
\end{thebibliography}
\end{document}